\documentclass[journal,twoside,A4paper]{IEEEtran}

\usepackage{cite}
\usepackage[pdftex]{graphicx}
\DeclareGraphicsExtensions{.pdf,.eps}
\usepackage{amsmath}
\usepackage{amssymb}
\usepackage{caption}
\usepackage{subcaption}
\usepackage{hyperref}
\hypersetup{
pdfpagemode=UseNone,
pdfstartview=FitH,
pdftitle=Simple Thermal Noise Estimation of Switched Capacitor Circuits Based on OTAs -- Part I: Amplifiers with Capacitive Feedback,
pdfauthor=Christian Enz,
pdfsubject=Noise in SC circuits,
pdfcreator=Christian Enz}
\usepackage[compress,capitalize]{cleveref}
\crefname{figure}{Fig.}{Figs.}
\crefname{equation}{}{}
\Crefname{equation}{Eq.}{Eqs.}
\crefname{secinapp}{Appendix}{appendices}
\Crefname{secinapp}{Appendix}{Appendices}

\newcommand{\super}[1]{\ensuremath{^{\mathrm{#1}}}}
\newcommand{\sub}[1]{\ensuremath{_{\mathrm{#1}}}}

\newcommand{\ELDO}{\textsc{{ELDO}\super{\copyright}}}

\newcommand{\TA}{transconductance amplifier} %
\newcommand{\TAs}{transconductance amplifiers} %
\newcommand{\TaH}{track \& hold} %

\newcommand{\Gm}{\ensuremath{G_m}}
\newcommand{\Gon}{\ensuremath{G_{on}}}
\newcommand{\Ron}{\ensuremath{R_{on}}}

\newcommand{\kT}{\ensuremath{k_B T}}

\newcommand{\hfb}{\ensuremath{h_{fb}}}

\newcommand{\ph}[1]{\ensuremath{\Phi_#1}}
\newcommand{\phase}[1]{phase \ensuremath{\Phi_#1}}
\newcommand{\Phase}[1]{Phase \ensuremath{\Phi_#1}}

\newcommand{\Var}[2]{\ensuremath{#1_{n#2}^2}}
\newcommand{\Varp}[3]{\ensuremath{\left. #1_{n#2}^2 \right|_{\Phi_#3}}}

\newcommand{\Cinf}{\ensuremath{C_{\infty}}}
\newcommand{\Cinfn}[1]{\ensuremath{C_{\infty (#1)}}}
\newcommand{\Cinfp}{\ensuremath{C_{\infty}'}}
\newcommand{\Cinfpn}[1]{\ensuremath{C_{\infty (#1)}'}}
\newcommand{\Co}{\ensuremath{C_{0}}}
\newcommand{\Con}[1]{\ensuremath{C_{0 (#1)}}}

\newcommand{\Cin}{\ensuremath{C_{in}}}
\newcommand{\Cout}{\ensuremath{C_{out}}}
\newcommand{\CL}{\ensuremath{C_L}}
\newcommand{\Ceq}{\ensuremath{C_{eq}}}

\newcommand{\ain}{\ensuremath{\alpha_{in}}}
\newcommand{\aL}{\ensuremath{\alpha_L}}

\newcommand{\Av}{\ensuremath{A_v}}
\newcommand{\aAv}{\ensuremath{|A_v|}}

\newcommand{\Ts}{\ensuremath{T_s}}
\newcommand{\tset}{\ensuremath{t_{set}}}

\newcommand{\bota}{\ensuremath{\beta_{ota}}}
\newcommand{\bswi}{\ensuremath{\beta_{sw}}}
\newcommand{\botap}[1]{\ensuremath{\left.\beta_{ota} \right|_{\Phi_#1}}}
\newcommand{\bswip}[1]{\ensuremath{\left. \beta_{sw} \right|_{\Phi_#1}}}

\newcommand{\num}{\ensuremath{n_2 s^2 + n_1 s + n_0}}
\newcommand{\den}{\ensuremath{d_3 s^3 + d_2 s^2 + d_1 s + d_0}} 

\begin{document}

\title{Simple Thermal Noise Estimation of Switched Capacitor Circuits Based on OTAs -- Part I: Amplifiers with Capacitive Feedback}

\author{\IEEEauthorblockN{
Christian Enz~\IEEEmembership{Fellow,~IEEE},
Antonino Caizzone~\IEEEmembership{Member,~IEEE},
Assim Boukhayma~\IEEEmembership{Member,~IEEE},\\
and
Fran\c{c}ois Krummenacher}

%\thanks{Manuscript submitted August 15, 2019.}
\thanks{Corresponding author: C. Enz (email: christian.enz@epfl.ch).}
\thanks{C. Enz, A. Caizzone and A. Boukhayma are with the Integrated Circuits Lab (ICLAB), Micro-engineering Institute, School of Engineering, EPFL.}
\thanks{F. Krummenacher is with the Electrical Engineering Institute, School of Engineering, EPFL.}}

\markboth{Part I: Amplifiers with Capacitive Feedback}%
{Enz \MakeLowercase{\textit{et al.}}: Simple Thermal Noise Estimation of SC Circuits Based on OTAs}

\maketitle

\begin{abstract}
This paper presents a simple method for estimating the thermal noise voltage variance in passive and active switched-capacitor (SC) circuits using operational transconductance amplifiers (OTA). The proposed method is based on the Bode theorem for passive network which is extended to active circuits based on OTAs with capacitive feedback. It allows for a precise estimation of the thermal noise voltage variance by simple inspection of three equivalent circuits avoiding the calculation of any transfer functions nor integrals. In this Part~I, the method is applied to SC amplifiers and \TaH{} circuits and successfully validated by means of transient noise simulations. Part~II extends the application of the method to integrators and active SC filters.
\end{abstract}


\IEEEpeerreviewmaketitle
	
\section{Introduction}\label{sec:introduction}

\IEEEPARstart{S}{witched-capacitor} (SC) circuits were invented in the 70's as a way to perform analog signal processing on-chip using the capacitors, switches and amplifiers available in MOS technologies \cite{bib:poschenrieder:1966, bib:fried:jssc:aug:1972}. They take advantage of the fact that the circuit characteristic only depends on capacitance ratios which turn out to be very accurate tanks to the excellent matching of capacitors. Additionally the frequency response of SC filters can be tuned by changing the clock frequency \cite{bib:gregorian:book:1986}. SC circuits have then been used broadly for various circuits including analog-to-digital and digital-to-analog converters \cite{bib:mccreary:jssc:dec:1972}. They are increasingly used in many more applications like radio frequency (RF) circuits \cite{bib:darvishi:jssc:dec:2013, bib:ghaffari:jssc:may:2011, bib:xu:jssc:may:2016} or sensor front-end circuits \cite{bib:boukhayma:jssc:sept:2016, bib:boukhayma:sensor:march:2016, bib:boukhayma:cms:icnf:2015} to perform various analog signal processing operations such as sampling, amplification or filtering.

The analysis of SC circuits has received considerable attention in the 80s' in particular for the computation of the noise \cite{bib:liou:tcas:april:1979, bib:vandewalle:ijct:jan:1981, bib:fischer:jssc:aug:1982, bib:gobet:tcas:jan:1983, bib:goette:tcas:april:1989, bib:toth:iscas:1991, bib:toth:tcas1:march:1999}. With the application of SC circuits to a wider range of analog and RF circuits, new computation techniques have also been proposed more recently \cite{bib:oliaei:tcas1:july:2003, bib:vasudevan:tcas1:feb:2004, bib:vasudevan:tcas1:nov:2004}. Today, modern circuit simulators allow to compute noise for example in the time domain using transient noise analysis \cite{bib:bolcato:iscas:1992}. However all of these techniques remain complex and are mostly focused on an efficient numerical computation of the PSD by dedicated EDA tools. They cannot be used for the derivation of simple analytical expressions of the noise voltage variance.

The performance of SC circuits is ultimately limited by the thermal and flicker noise (or $1/f$ noise) generated by the amplifiers and by the thermal noise coming from the switches. Since SC circuits are sampled-data systems, the broadband thermal noise is aliased into the Nyquist band, resulting in an increase of the noise power spectral density (PSD) by a factor equal to the ratio of the equivalent noise bandwidth to the Nyquist frequency which is usually much larger than one \cite{bib:gobet:elletters:sept:1981, bib:gobet:tcas:jan:1983, bib:enz:pieee:nov:1996}. The $1/f$ noise contribution can therefore usually be neglected and if it still remains important, the amplifier $1/f$ noise and offset can be reduced by increasing the transistor gate areas or eventually eliminated thanks to circuit techniques like auto-zeroing \cite{yen:jssc:dec:1983, bib:krummenacher:jssc:june:1982, bib:enz:el:20:23:1984, bib:enz:pieee:nov:1996} or chopper stabilization \cite{bib:hsieh:jssc:dec:1981, bib:enz:jssc:june:1987, bib:enz:pieee:nov:1996}. Under such conditions, the sampled thermal noise remains the dominant noise source particularly when minimal capacitance values are used, since the sampled noise voltage variance is inversely proportional to the capacitance.

Since the power consumption and silicon area are proportional to the capacitance \cite{bib:castello:tcas:sept:1985}, whereas the noise voltage variance is inversely proportional to the capacitance, it is crucial to identify which capacitances are setting the noise voltage variance. Unfortunately, the derivation of the noise PSD and variance is not easy because SC circuits are periodically time-varying circuits. The noise is therefore cyclostationary and usually characterized by the power spectral density (PSD) averaged over one period \cite{bib:gardner:mcgrawhill:book:1989}.

The optimization of SC circuits for achieving at the same time low-noise operation at low-power requires an accurate estimation of the noise variance. The latter is traditionally calculated for each phase in the frequency domain by first evaluating the transfer functions from all the noise sources to the node where the noise needs to be evaluated. The total noise PSD is then integrated over frequency to provide the noise variance. This approach is however quite tedious and impractical \cite{bib:schreier:tcas1:nov:2005} and for large networks, it becomes extremely cumbersome to get an analytical expressions \cite{bib:dastgheib:tcas:nov:2008}.

This work proposes a simple method for estimating the thermal noise voltage variance at any port of passive and active circuits made of operational transconductance amplifiers (OTA) with capacitive feedback as found in SC circuits \cite{bib:enz:icnf:2015}. The proposed method, based on the Bode theorem \cite{bib:bode:book:1945}, allows the calculation of the thermal noise voltage variance across any capacitor by simple inspection of several equivalent schematics made of capacitors only, avoiding the evaluation of complex transfer functions and
cumbersome integrals.

Part~I is dedicated to the derivation of the extended Bode theorem and its application to SC amplifiers and \TaH{} circuits. Part~II is focused on the application of the extended Bode theorem to SC filters. \Cref{sec:the bode theorem} of this first Part starts by recalling the Bode theorem, which is at the heart of the proposed method. \Cref{sec:the bode theorem_extension} then presents an extension of this theorem to OTAs with capacitive feedback as found in SC circuits. The method is then illustrated in \Cref{sec:verification on practical examples} by two simple examples, namely a SC amplifier and a SC \TaH{} circuit. The calculated noise in each case is compared to transient noise simulations results showing an excellent match. Conclusions are then given in \Cref{sec:conclusion}

\section{The Bode Theorem for Passive Networks}\label{sec:the bode theorem}

The Bode theorem \cite{bib:bode:book:1945,bib:weinrichter:iscas:1982,bib:furrer:thesis:1983} is a very efficient method to calculate the noise voltage variance at any port of an $RLC$ circuit and particularly of capacitive networks. However, this method is limited to passive $RLC$ networks.

\begin{figure}
  \centering
  \begin{subfigure}[t]{1\columnwidth}
    \centering
	\includegraphics[scale=0.8]{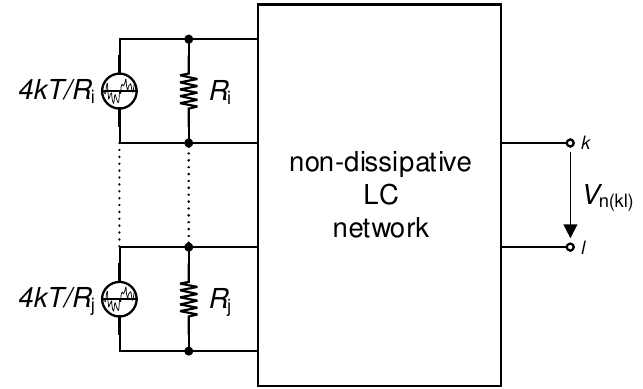}
	\caption{Thermal noise voltage variance in a passive $RLC$ network.}
	\label{fig:Bode_theorem_Vnkl}
  \end{subfigure}%
  \\
  \begin{subfigure}[t]{1\columnwidth}
    \centering
	\includegraphics[scale=0.8]{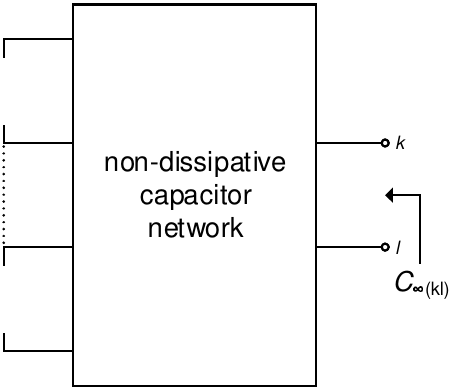}
	\caption{Circuit for calculating \Cinf{}.}
	\label{fig:Bode_theorem_Cinfkl}
  \end{subfigure}
  \\
  \begin{subfigure}[t]{1\columnwidth}
    \centering
	\includegraphics[scale=0.8]{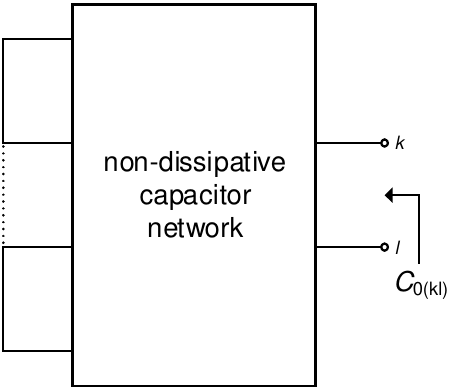}
	\caption{Circuit for calculating \Co{}.}
	\label{fig:Bode_theorem_C0kl}
  \end{subfigure}
  \label{fig:Bode_theorem}
\end{figure}

In linear circuits, the noise analysis is traditionally performed by integrating the noise PSD. This requires the calculation of the transfer functions from each uncorrelated noise source to the node where the noise has to be evaluated (for example at the circuit output) and then adding the obtained uncorrelated contributions. In case of a passive $RLC$ network, the thermal noise is generated in the resistors while the rest of the circuit made of ideal capacitors and inductors is noiseless. The circuit can then be represented as shown in \Cref{fig:Bode_theorem_Vnkl} where all resistors are modeled by a noiseless resistor in parallel with a noisy current source with power spectral density $4 \kT/R$. The thermal noise variance between any nodes $k$ and $l$ of the passive $RLC$ circuit can be calculated using the Bode theorem without the need for computing any integral by simple inspection of two equivalent circuits. The thermal noise voltage variance at any port is simply given by \cite{bib:bode:book:1945,bib:weinrichter:iscas:1982,bib:furrer:thesis:1983}
\begin{equation}\label{eqn:V2n_Bode_passive}
  \Var{V}{} = \kT \cdot \left(\frac{1}{\Cinf{}} - \frac{1}{\Co{}} \right),
\end{equation}
where $k_B$ is the Boltzmann constant and $T$ the absolute temperature. Capacitance \Cinf{} is defined as
\begin{equation}\label{eqn:Cinf}
  \frac{1}{\Cinf{}} = \lim_{s\to+\infty} s Z(s),
\end{equation}
which corresponds to the capacitance obtained when looking into the port after having removed all resistances from the circuit (or set them to infinity) as illustrated in \Cref{fig:Bode_theorem_Cinfkl}. Capacitance \Co{} is defined as
\begin{equation}\label{eqn:C0}
  \frac{1}{\Co{}} = \lim_{s\to0} s Z(s),
\end{equation}
which corresponds to the capacitance obtained when looking into the port after having replaced all resistances by a short circuit (or set them to zero) as illustrated in \Cref{fig:Bode_theorem_C0kl}.

\begin{figure}[!t]
  \centering
  \includegraphics[width=0.8\columnwidth]{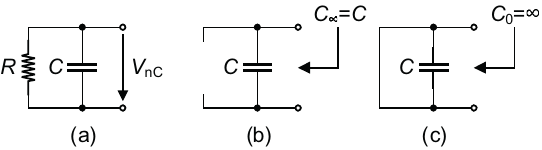}
  \caption{Example of a 1\super{st}-order low-pass passive filter.}
  \label{fig:First_order_RC}
\end{figure}

The simplest example of the application of the Bode theorem to a passive $RC$ circuit is the 1\super{st}-order low-pass filter illustrated in \Cref{fig:First_order_RC}a. As shown in \Cref{fig:First_order_RC}b, capacitance $\Cinf{}$ is obtained after removing the resistance and is therefore equal to $C$. Capacitance $\Co{}$ is obtained from \cref{eqn:C0} after replacing the resistance by a short resulting in $1/\Co{}=0$. Applying the Bode theorem \cref{eqn:V2n_Bode_passive}, the noise voltage variance is then simply equal to the well-known formula $\Var{V}{}=\kT/C$.

\section{Extension of the Bode Theorem to Transconductance Amplifiers with Capacitive Feedback}\label{sec:the bode theorem_extension}

\begin{figure}[!ht]
  \centering
  \begin{subfigure}[t]{0.49\columnwidth}
    \centering
    \includegraphics[scale=0.9]{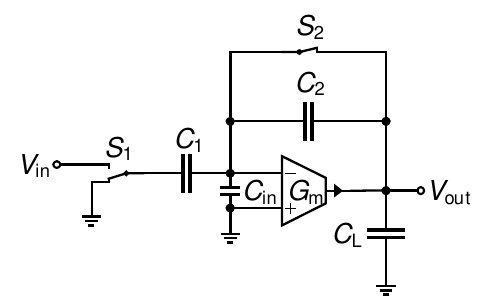}
    \caption{\centering \Phase{1}.}
    \label{fig:SC_amplifier_ph1}
  \end{subfigure}
  \begin{subfigure}[t]{0.49\columnwidth}
    \centering
    \includegraphics[scale=0.9]{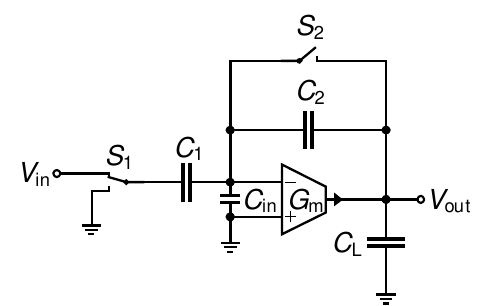}
    \caption{\centering \Phase{2}.}
    \label{fig:SC_amplifier_ph2}
  \end{subfigure}
  \caption{SC autozero (AZ) amplifier.}
  \label{fig:SC_amplifier}
\end{figure}
This Section presents an extension of the Bode theorem to be used for calculating the thermal noise voltage variance seen at any port of an OTA-based SC circuit during a given phase.

The Bode theorem presented above strictly applies only for the calculation of the thermal noise voltage variances of passive networks. However, it will be shown below that it can be extended to estimate the thermal noise voltage variance of amplifiers implemented with transconductance stages having a capacitive feedback. The simplest of such amplifier is the capacitive feedback amplifier shown in \Cref{fig:SC_amplifier}, where the amplifier is a differential OTA \cite{bib:gregorian:book:1986, bib:krummenacher:el:1:4:1981}. Note that the amplifier can also represent a single-ended transconductance stage as simple as a single transistor or cascoded transistor for improved dc gain \cite{bib:graymeyer:2009}. For more power-efficient implementation, the single-ended amplifier can be implemented by a simple inverter to take advantage of its current reuse feature \cite{bib:krummenacher:el:17:13:1981}. In the following discussion, the OTA is assumed to be ideal (infinite DC gain, offset free, no saturation) and can be modelled by a voltage controlled current source (VCCS).

The amplifier operates with two non overlapping phases: the autozero (AZ) \phase{1}, followed by the amplification \phase{2}. During \phase{1}, shown in \Cref{fig:SC_amplifier_ph1}, the amplifier output is connected to its input, discharging the feedback capacitor $C_2$. Assuming the DC gain of the OTA is infinite, capacitors $C_1$ and \Cin{} are then also discharged. In the amplification phase \ph{2}, switch $S_1$ is connected to the input, amplifying the input voltage. The voltage gain between the input and output voltage is simply equal to $A_v=-C_1/C_2$. Note that this amplifier is called autozero amplifier because it strongly reduces the OTA offset and flicker noise \cite{bib:enz:pieee:nov:1996}. It can be shown that the residual input-referred offset of the amplifier is equal to the OTA original offset divided by the amplifier voltage gain $A_v$ (assuming again that the OTA has an infinite DC gain). It can also be shown that the amplifier equivalent input-referred noise is free from the original OTA flicker noise and increased due to the aliasing of the broadband thermal noise due to the sampling process \cite{bib:enz:pieee:nov:1996}.

\begin{figure}[!t]
  \centering
  \begin{subfigure}[t]{1\columnwidth}
    \centering
    \includegraphics[scale=0.9]{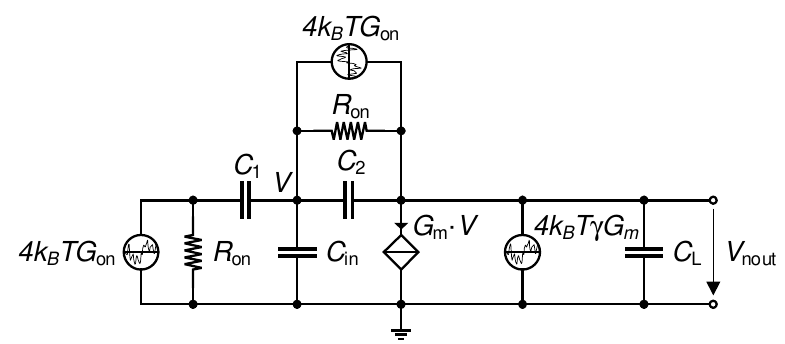}
    \caption{\centering \Phase{1}.}
    \label{fig:SC_amplifier_noise_ph1}
  \end{subfigure}
  \\
  \begin{subfigure}[t]{1\columnwidth}
    \centering
    \includegraphics[scale=0.9]{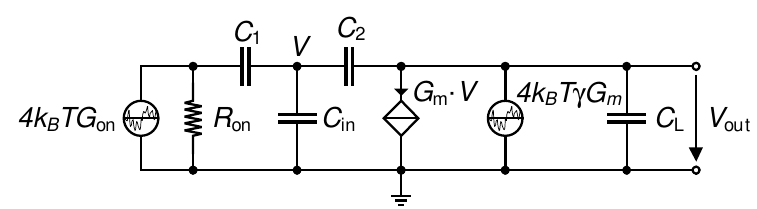}
    \caption{\centering \Phase{2}.}
    \label{fig:SC_amplifier_noise_ph2}
  \end{subfigure}
  \caption{Small-signal equivalent circuit of \Cref{fig:SC_amplifier} for the calculation of the output noise voltage.}
  \label{fig:SC_amplifier_noise}
\end{figure}

\begin{figure}[!t]
  \centering
  \begin{subfigure}[t]{1\columnwidth}
    \centering
    \includegraphics[scale=0.9]{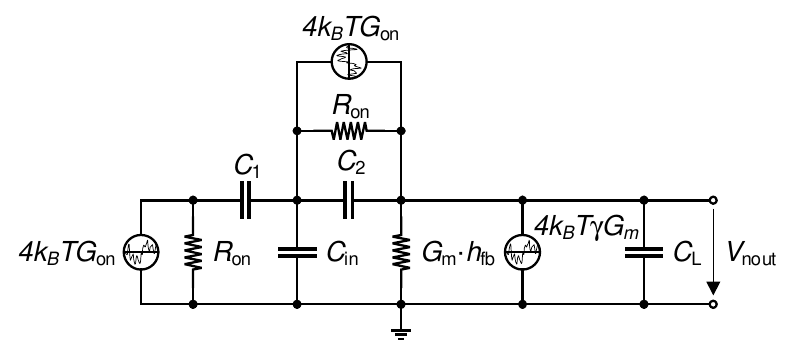}
    \caption{\centering \Phase{1}.}
    \label{fig:SC_amplifier_simplified_ph1}
  \end{subfigure}
  \\
  \begin{subfigure}[t]{1\columnwidth}
    \centering
    \includegraphics[scale=0.9]{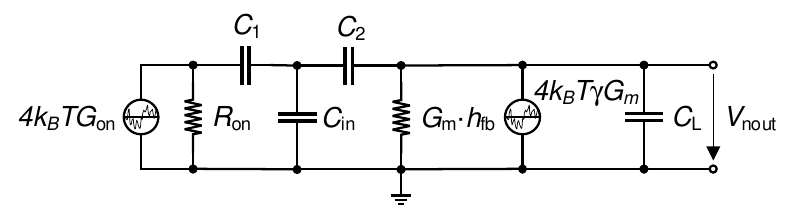}
    \caption{\centering \Phase{2}.}
    \label{fig:SC_amplifier_simplified_ph2}
  \end{subfigure}
  \caption{Equivalent circuit replacing the VCCS of \Cref{fig:SC_amplifier_noise} by a resistance with conductance $\Gm \cdot \hfb$ for the calculation of the output thermal noise voltage.}
  \label{fig:SC_amplifier_simplified}
\end{figure}

We are interested in the output noise variance at the end of the amplification phase \ph{2}, when the signal is actually read at the output. As mentioned above, thanks to the AZ process, the flicker noise of the OTA is strongly reduced \cite{bib:enz:pieee:nov:1996} and hence the noise at the amplifier output during phase \ph{2} is dominated by the thermal noise coming from the OTA and from the switches. The noise voltage variance at the output can be calculated in a classical way by integrating the output noise voltage PSD over frequency or calculating the equivalent noise bandwidth. An additional noise component needs to be accounted for at the end of phase \ph{2}, namely the noise that is generated across capacitor $C_1$ during phase \ph{1}. This noise is sampled as a noise charge on $C_1$ when the switch S\sub{1} opens at the end of \phase{1} and then transferred to the feedback capacitor $C_2$ during \phase{2}. The variance of this output noise voltage is obtained by first calculating the variance of the noise voltage across $C_1$ during \phase{1} by calculating the noise voltage PSD across $C_1$ and integrating it over frequency. This noise voltage variance corresponds to a frozen charge that is then transferred to the feedback capacitor during \phase{2}. Assuming that the OTA DC gain is infinite, the voltage across $C_2$ is equal to the output voltage. The output noise voltage variance due to the noise sampled on $C_1$ at the end of \phase{1} is simply equal to the noise voltage variance across $C_1$ multiplied by the square of the voltage gain \Av{}. Now, although the procedure is straightforward, it is actually not always possible to get simple analytical expressions for the noise voltage variances mentioned above. Of course the latter can always be calculated numerically using a simple \verb".NOISE" simulation, but for circuit optimization it is useful to have analytical expressions showing the dependence of the noise to the various components. The Bode theorem can unfortunately not be used because we now have an active component. However, we will show below that the noise voltage variances can be estimated by extending the original Bode theorem to circuits including amplifiers implemented as transconductance amplifiers with a capacitive feedback.

For the noise analysis, it is reasonable to use a small-signal analysis. The small-signal equivalent circuit of the amplifier of \Cref{fig:SC_amplifier} used to calculate the output noise voltage is shown in \Cref{fig:SC_amplifier_noise}, where the noise current sources across resistances \Ron{} represent the thermal noise of the switches having a PSD $4 \kT \Gon$, where $\Gon=1/\Ron$, whereas the noise current source across the VCCS represents the OTA thermal noise referred to the output with a PSD $4 \kT \gamma \Gm$, where $\gamma$ is the thermal noise excess factor close to unity for a single transistor and usually larger than 2 for a differential OTA.

In both phases \ph{1} and \ph{2}, neglecting the current noise sources, the voltage at the transconductance amplifier virtual ground is only a function of the output voltage
\begin{equation}\label{eqn:v}
  V = \hfb \cdot V_{out},
\end{equation}
where $h_{fb} \triangleq V/V_{out}$ is the feedback voltage gain. During \phase{1} it is simply equal to unity, whereas during \phase{2} it is given by
\begin{equation}\label{eqn:hfb}
  \hfb = \frac{C_2}{C_1+C_2+\Cin}.
\end{equation}

\begin{figure}
  \centering
  \begin{subfigure}[t]{1\columnwidth}
    \centering
	\includegraphics[scale=0.8]{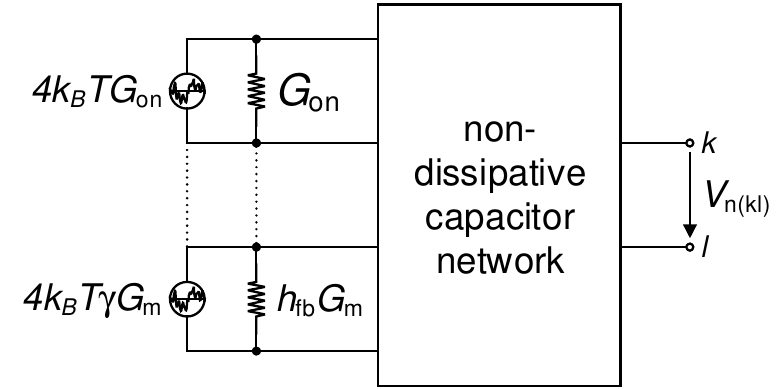}
	\caption{\TA-based SC circuit in one phase represented as a passive $RC$ network.}
	\label{fig:Bode_extension_Vnkl}
  \end{subfigure}%
  \\
  \begin{subfigure}[t]{1\columnwidth}
    \centering
	\includegraphics[scale=0.8]{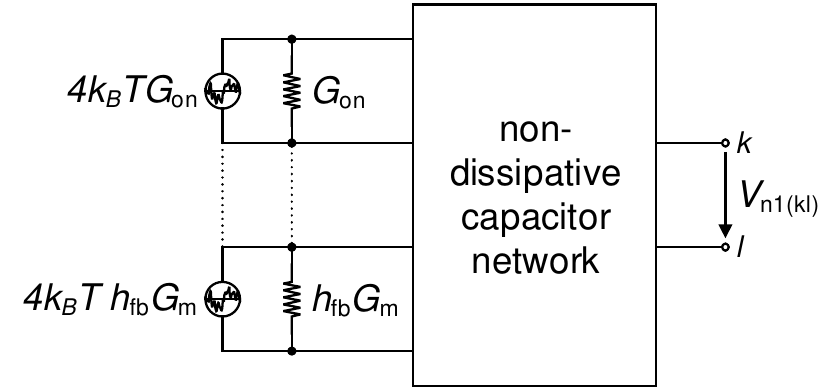}
	\caption{Circuit of \cref{fig:Bode_extension_Vnkl} with all conductances having the same noise temperature $T$.}
	\label{fig:Bode_extension_Vn1kl}
  \end{subfigure}
  \\
  \begin{subfigure}[t]{1\columnwidth}
    \centering
	\includegraphics[scale=0.8]{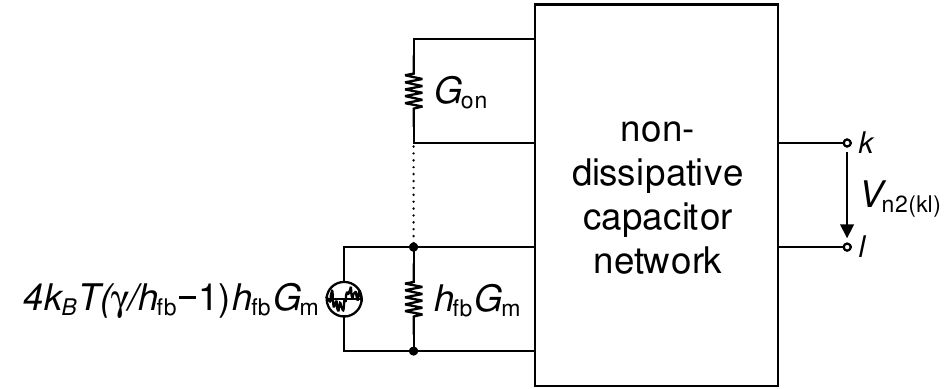}
	\caption{Circuit of \cref{fig:Bode_extension_Vnkl} without the switch noise sources and accounting for the excess noise of the \TA{} with a noise temperature $(\gamma/\hfb - 1)\cdot T$.}
	\label{fig:Bode_extension_Vn2kl}
  \end{subfigure}
  \\
  \begin{subfigure}[t]{1\columnwidth}
    \centering
	\includegraphics[scale=0.8]{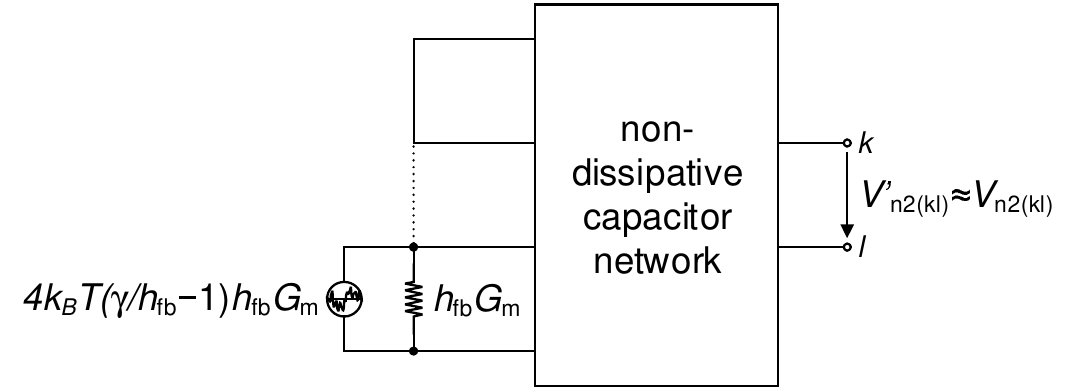}
	\caption{Simplification of the circuit of \cref{fig:Bode_extension_Vn2kl} accounting for the fact that usually $G_{on} \gg \hfb \cdot \Gm$.}
	\label{fig:Bode_extension_Vn2kl_prime}
  \end{subfigure}
  \caption{Simplified schematic of \TAs-based circuit \cite{bib:enz:icnf:2015}.}
  \label{ext_bode_Vn}
\end{figure}

The circuits of \Cref{fig:SC_amplifier_noise} can therefore be simplified by replacing the VCCS by a simple resistance having a conductance equal to $\Gm \cdot \hfb$ resulting in the simplified circuits shown in \Cref{fig:SC_amplifier_simplified}. The later circuits can now be considered as passive and can be represented as in \Cref{fig:Bode_extension_Vnkl}. However, the Bode theorem cannot be applied directly because the noise current source corresponding to the conductance $\hfb \cdot \Gm$ is not equal to $4 \kT \hfb \Gm$ like for the noise sources associated to the switch resistances. In order to apply the Bode theorem, the OTA noise current source PSD $4 \kT \gamma \Gm$ can be split into the sum of $4 \kT \hfb \Gm$ and a term $4 \kT (\gamma-\hfb) T \Gm$ that includes the OTA thermal noise excess. The circuit of \Cref{fig:Bode_extension_Vnkl} can hence be decomposed into two circuits, the circuit shown in \cref{fig:Bode_extension_Vn1kl}, where all the conductances have the same noise temperature $T$, and the circuit of \cref{fig:Bode_extension_Vn2kl}, where the switches are considered noiseless and the noise source corresponding to conductance $\hfb \cdot \Gm$ is considered to have a noise temperature equal to $(\gamma/\hfb - 1)\cdot T$. The variance of the noise voltage $V_{n(kl)}^2$ between any node $k$ and $l$ of the circuit of \cref{fig:Bode_extension_Vnkl} can then be calculated using the superposition of the noise sources as the sum of the noise voltage variance $V_{n1(kl)}^2$ of the circuit shown in \cref{fig:Bode_extension_Vn1kl}, where all the conductances have the same noise temperature $T$, and noise voltage variance $V_{n2(kl)}^2$ of the circuit shown in \cref{fig:Bode_extension_Vn2kl}, corresponding to the excess noise in the equivalent conductance of the OTA with a noise temperature $(\gamma/\hfb - 1)\cdot T$
\begin{equation}\label{v_ncki}
  \Var{V}{(kl)} = \Var{V}{1(kl)} + \Var{V}{2(kl)}.
\end{equation}

\begin{figure*}[!hbt]
  \centering
  \begin{subfigure}[t]{.33\textwidth}
    \centering
	\includegraphics[scale=.8]{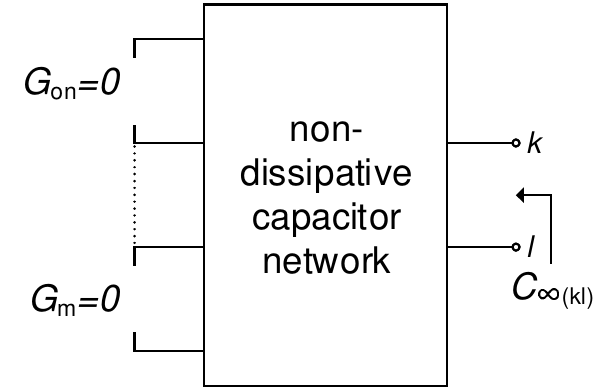}
	\caption{Equivalent circuit used for the calculation of \Cinfn{kl}: all switches and OTAs of the SC circuit are removed.}
	\label{fig:Equivalent_circuit_Cinf}
  \end{subfigure}%
  \begin{subfigure}[t]{.33\textwidth}
    \centering
	\includegraphics[scale=.8]{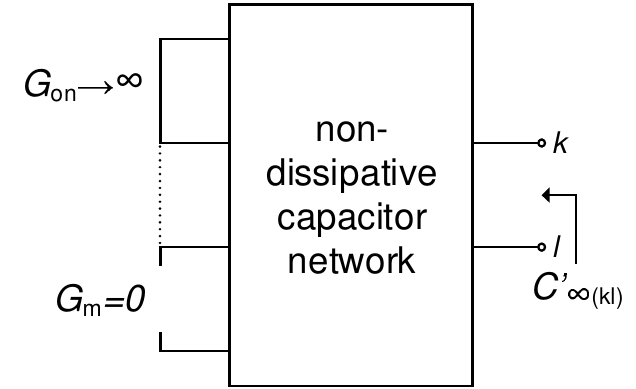}
	\caption{Equivalent circuit used for the calculation of \Cinfpn{kl}: all switches that are closed during the clock phase in consideration are replaced by short-circuits and all OTAs of the SC circuit	are removed.}
	\label{fig:Equivalent_circuit_Cinf_prime}
  \end{subfigure}
  \begin{subfigure}[t]{.33\textwidth}
    \centering
	\includegraphics[scale=.8]{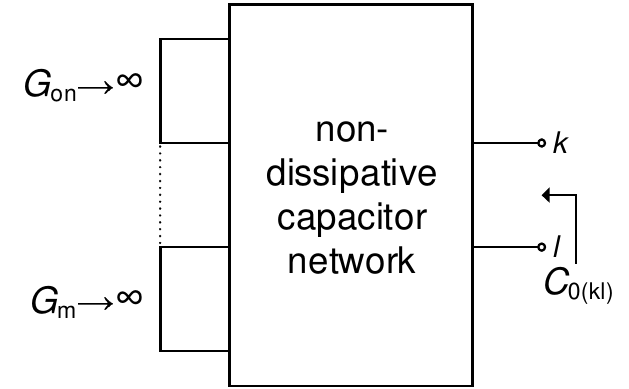}
	\caption{Equivalent circuit used for the calculation of \Con{kl}: all switches that are closed during the clock phase in consideration are replaced by short-circuits and all OTAs of the SC circuit have their output shorted to ground.}
	\label{fig:Equivalent_circuit_C0}
  \end{subfigure}
  \caption{Capacitances calculation for the extended Bode theorem \cite{bib:enz:icnf:2015}.}
  \label{fig:Equivalent_circuit_C}
\end{figure*}

The Bode theorem for passive networks can then be applied to the circuit shown in \Cref{fig:Bode_extension_Vn1kl} to calculate the noise voltage variance $V_{n1(kl)}^2$ as
\begin{equation}\label{eqn:V2n1}
  \Var{V}{1(kl)} = \kT \cdot \left(\frac{1}{\Cinfn{kl}} - \frac{1}{\Con{kl}} \right),
\end{equation}
where \Cinfn{kl} corresponds to the capacitance seen when looking between the nodes $k$ and $l$ when all the switches and \TAs{} are removed, and \Con{kl} corresponds to the capacitance obtained looking between the nodes $k$ and $l$ when the switches are replaced by short-circuits and all OTAs have their output shorted to ground.

The circuit of \Cref{fig:Bode_extension_Vn2kl} can be further simplified considering that usually $\Gon \gg \hfb \cdot \Gm$ and hence the on-conductances of the switches can be replaced by short-circuits resulting in the circuit shown in \cref{fig:Bode_extension_Vn2kl_prime}. The noise voltage variance $V_{n2(kl)}^2$ accounting for the OTA excess noise temperature can therefore be estimated by applying the Bode theorem to the circuit of \Cref{fig:Bode_extension_Vn2kl_prime} resulting in
\begin{equation}\label{eqn:V2n2}
  \Var{V}{2(kl)} \cong \kT \cdot \left(\frac{\gamma}{\hfb} - 1 \right) \cdot \left(\frac{1}{\Cinfpn{kl}} - \frac{1}{\Con{kl}} \right),
\end{equation}
where \Cinfpn{(kl)} corresponds to the capacitance seen between the nodes $k$ and $l$ when all the switches are replaced by short-circuits and the \TAs{} are removed.

The total thermal noise voltage variance seen between nodes $k$ and $l$ is then given by summing \cref{eqn:V2n1} and \cref{eqn:V2n2}, resulting in
\begin{equation}\label{eqn:V2n_Bode_OTA}
  \Var{V}{(kl)} = \kT \cdot \left[\frac{1}{\Cinfn{kl}} + \frac{\gamma/\hfb - 1}{\Cinfpn{kl}}
  - \frac{\gamma/\hfb}{\Con{kl}}\right].
\end{equation}

\Cref{eqn:V2n_Bode_OTA} is central to this calculation method. It shows that the computation of $\Var{V}{(kl)}$, for example at the amplifier output, only requires the evaluation of the three capacitances \Cinfn{kl}, \Cinfpn{kl} and \Con{kl}. The latter can easily be calculated by inspection of the three equivalent circuits depicted in \Cref{fig:Equivalent_circuit_C} which are each composed only of capacitors.

The extension of the Bode theorem presented in this Section will now be illustrated and validated by transient noise simulations for various SC circuits in the next Section.

\section{Practical examples of thermal noise estimation in OTA-based SC circuits}\label{sec:verification on practical examples}
\subsection{SC Amplifier}
\subsubsection{Analysis}
\begin{figure*}[!htb]
  \centering
  \begin{subfigure}[t]{.24\textwidth}
    \centering
	\includegraphics[scale=.8]{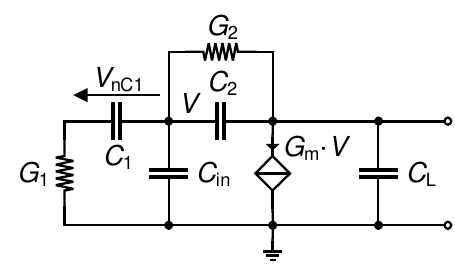}
	\caption{\centering Equivalent SC amplifier circuit during \phase{1}.}
	\label{fig:SC_amplifier_ph1a}
  \end{subfigure}
  \begin{subfigure}[t]{.24\textwidth}
    \centering
    \includegraphics[scale=.8]{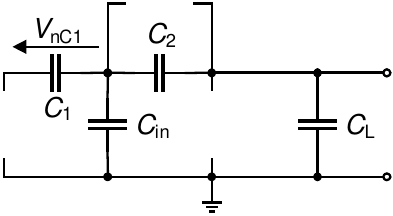}
    \caption{\centering SC amplifier circuit for calculation of \Cinfn{kl}:\\
    $\Cinfn{C_1}=C_1$.}
    \label{fig:SC_Tamplifier_ph1b}
  \end{subfigure}
  \begin{subfigure}[t]{.24\textwidth}
    \centering
	\includegraphics[scale=.8]{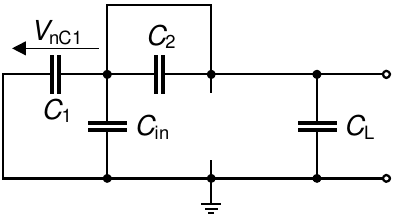}
	\caption{\centering SC amplifier circuit for calculation of \Cinfpn{kl}:\\
    $\Cinfpn{C_1}=C_1+\Cin+\CL$.}
	\label{fig:SC_amplifier_ph1c}
  \end{subfigure}
  \begin{subfigure}[t]{.24\textwidth}
    \centering
	\includegraphics[scale=.8]{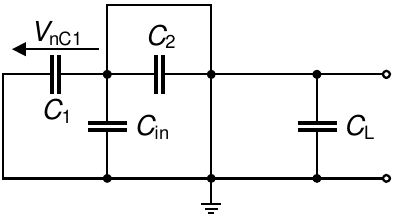}
	\caption{\centering SC amplifier circuit for calculation of \Con{kl}:\\
    $\Con{out}=\infty$.}
	\label{fig:SC_amplifier_ph1d}
  \end{subfigure}
  \caption{SC amplifier equivalent circuit schematics for \phase{1}.}
  \label{fig:SC_amplifier_ph1abcd}
\end{figure*}

Let's now get back to the SC amplifier shown in \Cref{fig:SC_amplifier} and apply the extended Bode theorem to calculate the noise voltage variances. We start calculating the noise voltage variance across the sampling capacitor $C_1$ during \phase{1} \Varp{V}{C_1}{1}. To this purpose, we need to calculate the three capacitances seen across $C_1$, namely \Cinfn{C_1}, \Cinfpn{C_1} and \Con{C_1} during \phase{1}. They can easily be calculated from the equivalent circuits shown in \Cref{fig:SC_amplifier_ph1abcd}, resulting in
\begin{subequations}
  \begin{align}
    \Cinfn{C_1} &= C_1,\label{eqn:SC_amplifier_Cinf_ph1}\\
    \Cinfpn{C_1} &= C_1 + \Cin + \CL,\label{eqn:SC_amplifier_Cinfp_ph1}\\
    \Con{C_1} &= \infty.\label{eqn:SC_amplifier_C0_ph1}
  \end{align}
\end{subequations}
Recognizing that the feedback gain during \phase{1} is simply equal to unity, the noise voltage variance across $C_1$ during \phase{1} can be evaluated from \cref{eqn:V2n_Bode_OTA} as
\begin{equation}\label{eqn:SC_amplifier_V2nC1_ph1}
  \Varp{V}{C_1}{1} = \frac{\kT}{C_1+\Cin+\CL} \cdot \left(\gamma + \frac{\Cin+\CL}{C_1}\right).
\end{equation}
\Cref{eqn:SC_amplifier_V2nC1_ph1} is actually identical to the result derived analytically in Appendix~\ref{sec:SC_amplifier_appendix} by calculating the noise contributions from each of the noise sources shown in \cref{fig:SC_amplifier_noise_ph1}, namely the two switches and the OTA. Each contribution is obtained by first calculating the transfer function from each noise source to the voltage across capacitor $C_1$ and integrating the corresponding noise PSD over frequency. The noise voltage variances due to each of the noise source assuming that the switch resistances are negligible compared to $1/\Gm$ (i.e. $\Gm \Ron \ll1$) are given in \Cref{tab:SC_amplifier_V2nC1p1i}. The total noise voltage variance is obtained by summing these three variances resulting in the result shown in the last row of \Cref{tab:SC_amplifier_V2nC1p1i} which is identical to \cref{eqn:SC_amplifier_V2nC1_ph1}. Note that when setting $\Cin=0$ in \cref{eqn:SC_amplifier_V2nC1_ph1}, we also get the same result than found in \cite{bib:caizzone:newcas:2018} which has been computed using the equivalent noise bandwidth approach. As expected, \cref{eqn:SC_amplifier_V2nC1_ph1} does not depend on $C_2$ since the later is short-circuited. This happens even though the individual contributions of each switch depends on $C_2$ as shown in \Cref{tab:SC_amplifier_V2nC1p1i}, but their sum does not depend on $C_2$ anymore.

To this noise voltage variance corresponds a noise charge sampled on $C_1$ at the end of \phase{1} and having a variance
\begin{equation}\label{eqn:SC_amplifier_QnC1_ph1}
  \Varp{Q}{C_1}{1} = C_1^2 \cdot \Varp{V}{C_1}{1}.
\end{equation}
This charge is then injected to the virtual ground at the beginning of \phase{2} and transferred to the feedback capacitor $C_2$ during \phase{2} thanks to the action of the OTA. Assuming that the latter has an infinite DC gain and a zero offset voltage, the output voltage is equal to the voltage across $C_2$. The output noise voltage variance due to the noise sampled on $C_1$ at the end of \phase{1} is then given by
\begin{equation}\label{eqn:SC_amplifier_V2nout_ph1_1}
  \begin{split}
    &\Varp{V}{out}{1} = \frac{\Varp{Q}{C_1}{1}}{C_2^2} = \left(\frac{C_1}{C_2}\right)^2 \cdot \Varp{V}{C_1}{1}\\
  &= \left(\frac{C_1}{C_2}\right)^2 \cdot \frac{\kT}{C_1+\Cin+\CL} \cdot \left(\gamma + \frac{\Cin+\CL}{C_1}\right).
  \end{split}
\end{equation}

\begin{figure*}[!htb]
  \centering
  \begin{subfigure}[t]{.24\textwidth}
    \centering
	\includegraphics[scale=.8]{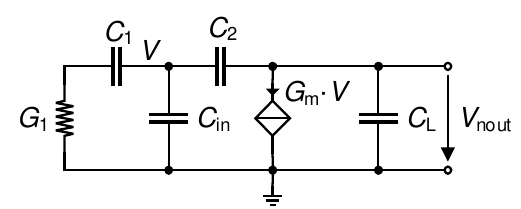}
	\caption{\centering Equivalent SC amplifier circuit during \phase{2}.}
	\label{fig:SC_amplifier_ph2a}
  \end{subfigure}
  \begin{subfigure}[t]{.24\textwidth}
    \centering
    \includegraphics[scale=.8]{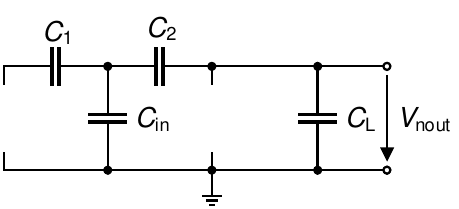}
    \caption{\centering SC amplifier circuit for calculation of \Cinfn{kl}:\\
    $\Cinfn{out}=\CL+\frac{C_2 \Cin}{C_2+\Cin}$.}
    \label{fig:SC_Tamplifier_ph2b}
  \end{subfigure}
  \begin{subfigure}[t]{.24\textwidth}
    \centering
	\includegraphics[scale=.8]{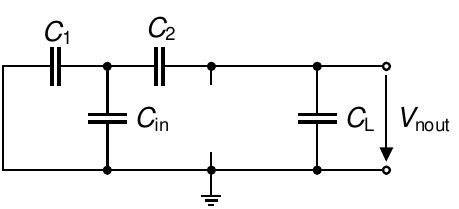}
	\caption{\centering SC amplifier circuit for calculation of \Cinfpn{kl}:\\
    $\Cinfpn{out}=\CL+\frac{C_2(C_1+\Cin)}{C_1+C_2+\Cin}$.}
	\label{fig:SC_amplifier_ph2c}
  \end{subfigure}
  \begin{subfigure}[t]{.24\textwidth}
    \centering
	\includegraphics[scale=.8]{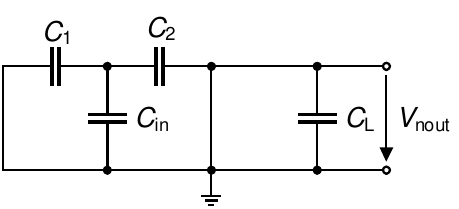}
	\caption{\centering SC amplifier circuit for calculation of \Con{kl}:\\
    $\Con{out}=\infty$.}
	\label{fig:SC_amplifier_ph2d}
  \end{subfigure}
  \caption{SC amplifier equivalent circuit schematics for \phase{2}.}
  \label{fig:SC_amplifier_ph2abcd}
\end{figure*}

The variance of the output noise voltage during \phase{2} is simply evaluated using \cref{eqn:V2n_Bode_OTA} which requires the calculation of the three capacitances \Cinfn{out}, \Cinfpn{out} and \Con{out} seen from the output. The later can be calculated from the equivalent schematics shown in \Cref{fig:SC_amplifier_ph2abcd} resulting in
\begin{subequations}
  \begin{align}
    \Cinfn{out} &= \CL + \frac{C_2 \Cin}{C_2+\Cin},\label{eqn:SC_amplifier_Cinf_ph2}\\
    \Cinfpn{out} &= \CL + \frac{C_2 (C_1+\Cin)}{C_1+C_2+\Cin},\label{eqn:SC_amplifier_Cinfp_ph2}\\
    \Con{out} &= \infty.\label{eqn:SC_amplifier_C0_ph2}
  \end{align}
\end{subequations}
\Cref{eqn:V2n_Bode_OTA} also requires the feedback gain $\hfb$ which during \phase{2} is given by
\begin{equation}\label{eqn:SC_amplifier_hfb_ph2}
  \left.\hfb\right|_{\Phi_2} = \frac{C_2}{C_1+C_2+\Cin}.
\end{equation}
The output noise voltage variance during \phase{2} can now be evaluated from \cref{eqn:V2n_Bode_OTA} leading to
\begin{equation}\label{eqn:SC_amplifier_V2nout_ph2}
  \Varp{V}{out}{2} = \frac{\kT}{C_2} \cdot (\gamma \cdot \botap{2} + \bswip{2}),
\end{equation}
where
\begin{subequations}
  \begin{align}
    \botap{2} &= \frac{(C_1 + C_2 + \Cin)^2}{B},\label{eqn:SC_amplifier_betaota_ph2_1}\\
    \bswip{2} &= \frac{C_1 C_2^3}{(C_2 \Cin + C_2 \CL + \Cin \CL) \cdot B},\label{eqn:SC_amplifier_betaswitch_ph2_1}
  \end{align}
\end{subequations}
with
\begin{equation}\label{eqn:SC_amplifier_B}
  B = C_1 C_2 + C_2 \Cin + C_1 \CL + C_2 \CL + \Cin \CL.
\end{equation}
Note that \cref{eqn:SC_amplifier_V2nout_ph2} is identical to the result derived in Appendix~\ref{sec:SC_amplifier_appendix} using the classical approach described above and given in the last row of \Cref{tab:SC_amplifier_V2noutp2i}.

\Cref{eqn:SC_amplifier_betaota_ph2_1} and \cref{eqn:SC_amplifier_betaswitch_ph2_1} can be rewritten as
\begin{subequations}
  \begin{align}
    \botap{2} &= \frac{(\aAv + \ain + 1)^2}{D},\label{eqn:SC_amplifier_betaota_ph2_2}\\
    \bswip{2} &= \frac{\aAv}{(\ain + \aL + \ain \aL) \cdot D},\label{eqn:SC_amplifier_betaswitch_ph2_2}\\
    D &= \frac{B}{C_2^2} = \aAv + \aL \cdot (\aAv+\ain+1) + \ain,\label{eqn:SC_amplifier_D}
  \end{align}
\end{subequations}
where $\aAv \triangleq C_1/C_2$, $\ain \triangleq \Cin/C_2$ and $\aL \triangleq \CL/C_2$.

Assuming that $\aAv \gg 1$, $\ain < 1$ and $\aL < 1$, $D \cong \aAv \cdot (\aL + 1)$ and equations \cref{eqn:SC_amplifier_betaota_ph2_2} and \cref{eqn:SC_amplifier_betaswitch_ph2_2} simplify to
\begin{subequations}
  \begin{align}
    \botap{2} &\cong \frac{\aAv}{\aL + 1},\label{eqn:SC_amplifier_betaota_ph2_approx}\\
    \bswip{2} &\cong \frac{1}{(\ain + \aL + \ain \aL)(\aL + 1)},\label{eqn:SC_amplifier_betaswitch_ph2_approx}
  \end{align}
\end{subequations}

Before calculating the total noise voltage variance at the output, we can rewrite \cref{eqn:SC_amplifier_V2nout_ph1_1} as
\begin{equation}\label{eqn:SC_amplifier_V2nout_ph1_2}
  \Varp{V}{out}{1} = \frac{\kT}{C_2} \cdot \left(\gamma \cdot \botap{1} + \bswip{1}\right),
\end{equation}
where
\begin{subequations}\label{eqn:SC_amplifier_beta_ph1}
  \begin{align}
    \botap{1} &= \frac{\aAv^2}{\aAv + \ain + \aL} \cong \aAv,\label{eqn:SC_amplifier_betaota_ph1}\\
    \bswip{1} &= \frac{\aAv \cdot (\ain + \aL)}{\aAv + \ain + \aL} \cong \ain + \aL.\label{eqn:SC_amplifier_betaswitch_ph1}
  \end{align}
\end{subequations}

The total output noise voltage variance at the end of \phase{2} is then given by summing \cref{eqn:SC_amplifier_V2nout_ph1_2} and \cref{eqn:SC_amplifier_V2nout_ph2}, resulting in
\begin{equation}\label{eqn:SC_amplifier_Vnouttot}
  \Var{V}{out} = \Varp{V}{out}{1} + \Varp{V}{out}{2} =  \frac{\kT}{C_2} \cdot (\gamma \cdot \bota + \bswi),
\end{equation}
where
\begin{subequations}\label{eqn:SC_amplifier_beta_tot}
  \begin{align}
    \bota &= \botap{1} + \botap{2} \cong \aAv \cdot \frac{\aL+2}{\aL+1},\label{eqn:SC_amplifier_betaota_tot}\\
    \begin{split}
      \bswi &= \bswip{1} + \bswip{2} \cong\\
      &\cong \ain+\aL+\frac{1}{(\ain+\aL+\ain \aL)(\aL+1)}.\label{eqn:SC_amplifier_betaswitch_tot}
    \end{split}
  \end{align}
\end{subequations}

It can be shown that the effect of $\Cin$ is negligible as long as the DC gain of the OTA is infinite. The switch contribution given by \cref{eqn:SC_amplifier_betaswitch_tot} can be further simplified by setting $\Cin=0$ (or $\ain=0$) resulting in
\begin{equation}\label{eqn:eqn:SC_amplifier_betaswitch_tot_approx}
  \bswi \cong \aL+\frac{1}{\aL(\aL+1)}.
\end{equation}

\subsubsection{Simulations}

The above results for the SC amplifier have been verified by transient noise simulation \cite{bib:bolcato:iscas:1992} in \ELDO. The simulations are performed on a circuit where the OTA is modeled by a simple VCCS and the switches are modelled by an ideal switch in series with a noisy resistor of resistance $\Ron=5\;k\Omega$. The noise of the OTA is generated by a noisy resistor of value $\gamma/\Gm$ and injected at the OTA output by means of a VCCS with a unity transconductance. The sampling period has been chosen equal to $1\;\mu s$ and the temperature is set to $T=300\;K$.

\begin{figure}[!t]
  \centering
  \includegraphics[width=1.0\columnwidth]{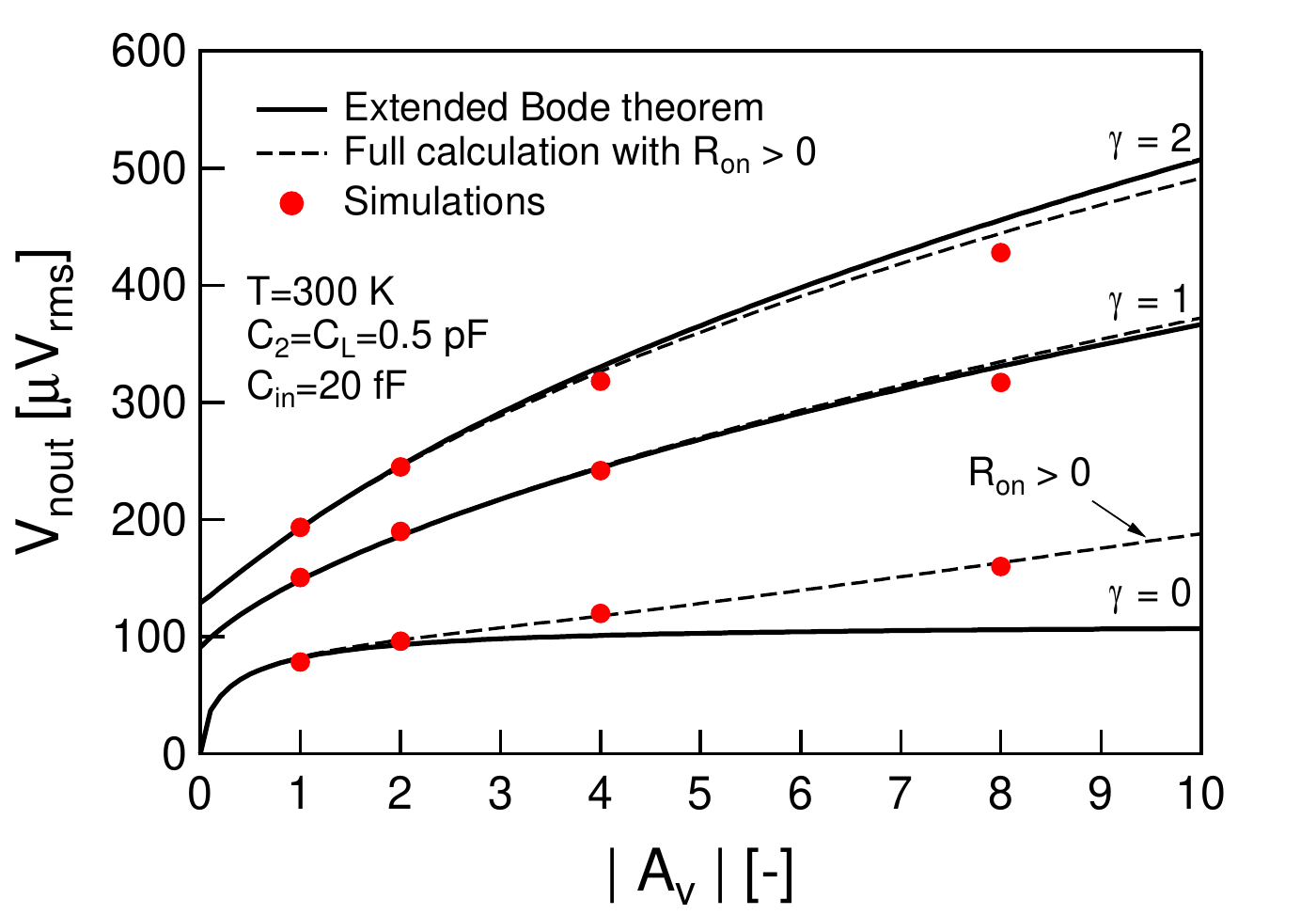}
  \caption{Output noise rms voltage versus the amplifier gain $\aAv$ for $\gamma=0,1,2$.}
  \label{fig:SC_amplifier_Vnout_vs_Av}
\end{figure}

\Cref{fig:SC_amplifier_Vnout_vs_Av} shows the output noise rms voltage versus the gain $\aAv$ for different values of the OTA excess noise factor $\gamma=0, 1, 2$. The simulations have been performed for different gains $\aAv=1, 2, 4, 8$ by changing the value of $C_1$ keeping $C_2=\CL=0.5\;pF$. When increasing capacitance $C_1$, it also increases the effective load capacitance $\Cout=\CL+(1-\beta)C_2$ with $\beta=C_2/(C_1+C_2+\Cin)$ and the settling time $\tset=\Ceq/\Gm$ where $\Ceq=\Cout/\beta$. The VCCS transconductance $\Gm$ has therefore been chosen to keep a constant settling time $\tset=\Ts/10=100\;ns$ for each values of $C_1$ and hence of $\aAv$. Note that the influence of the input capacitance $\Cin$ is negligible and the later has been set to a realistic value of $\Cin=20\;fF$. The simulation results for $\gamma=1$ and $\gamma=2$ are very close to the estimation computed from \cref{eqn:SC_amplifier_Vnouttot}. However, a small deviation is observed for the case where $\gamma=0$ which corresponds to the noise generated by the switches only. The simulation results are slightly larger than the values predicted by \cref{eqn:SC_amplifier_Vnouttot}, particularly for the maximum gain $\aAv=8$. The reason for this is that larger gains require larger $\Gm$ resulting in the product $\Gm \cdot \Ron$ increasing to about 0.27 which does no more fulfill the assumption of $\Gm \cdot \Ron \ll 1$ used in the Bode theorem derivation. The impact of a non-zero $\Gm \cdot \Ron$ has been checked using the full analytical expressions obtained from the classical analysis detailed in Appendix~\ref{sec:SC_amplifier_appendix}\footnote{Note that, for the sake of compactness,  the analytical expressions including the effect of a non-zero $\Gm \cdot \Ron$ could not be included in the Appendix~\ref{sec:SC_amplifier_appendix} because they are rather large expressions.}. The results are plotted in \Cref{fig:SC_amplifier_Vnout_vs_Av} by dashed lines which are very close to the simulation results, confirming the origin of the deviation.

\begin{figure}[!t]
  \centering
  \includegraphics[width=1.0\columnwidth]{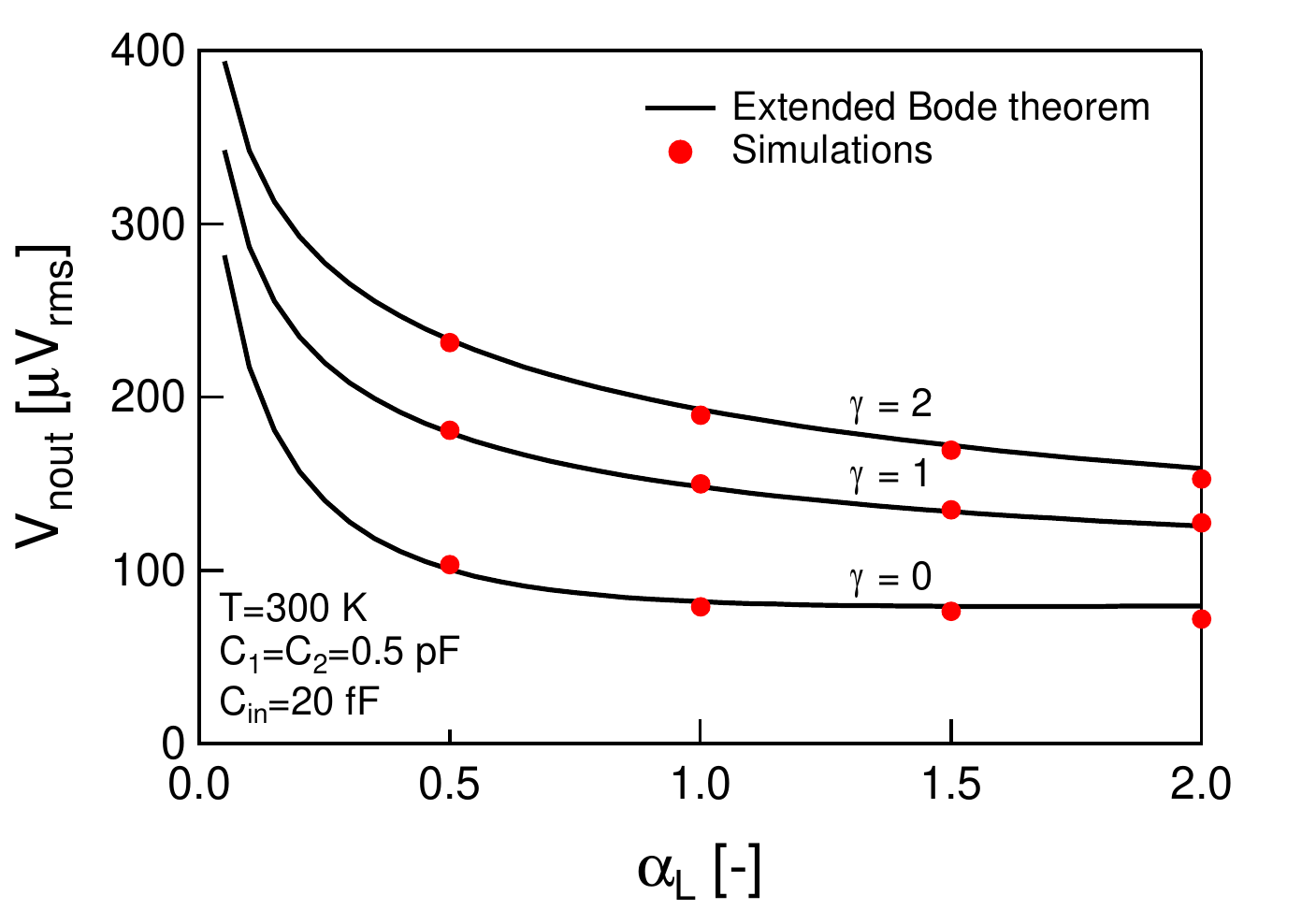}
  \caption{Output noise rms voltage versus $\aL$ for $\gamma=0,1,2$.}
  \label{fig:SC_amplifier_Vnout_vs_alphaL}
\end{figure}

\Cref{fig:SC_amplifier_Vnout_vs_alphaL} shows the output noise rms voltage versus $\aL$ for $\gamma=0,1,2$ and for a unity voltage gain $\aAv=1$ ($C_1=C_2=0.5\;pF$). The noise simulation results are very close to the estimation using \cref{eqn:SC_amplifier_Vnouttot}.

\begin{figure}[!t]
  \centering
  \includegraphics[width=1.0\columnwidth]{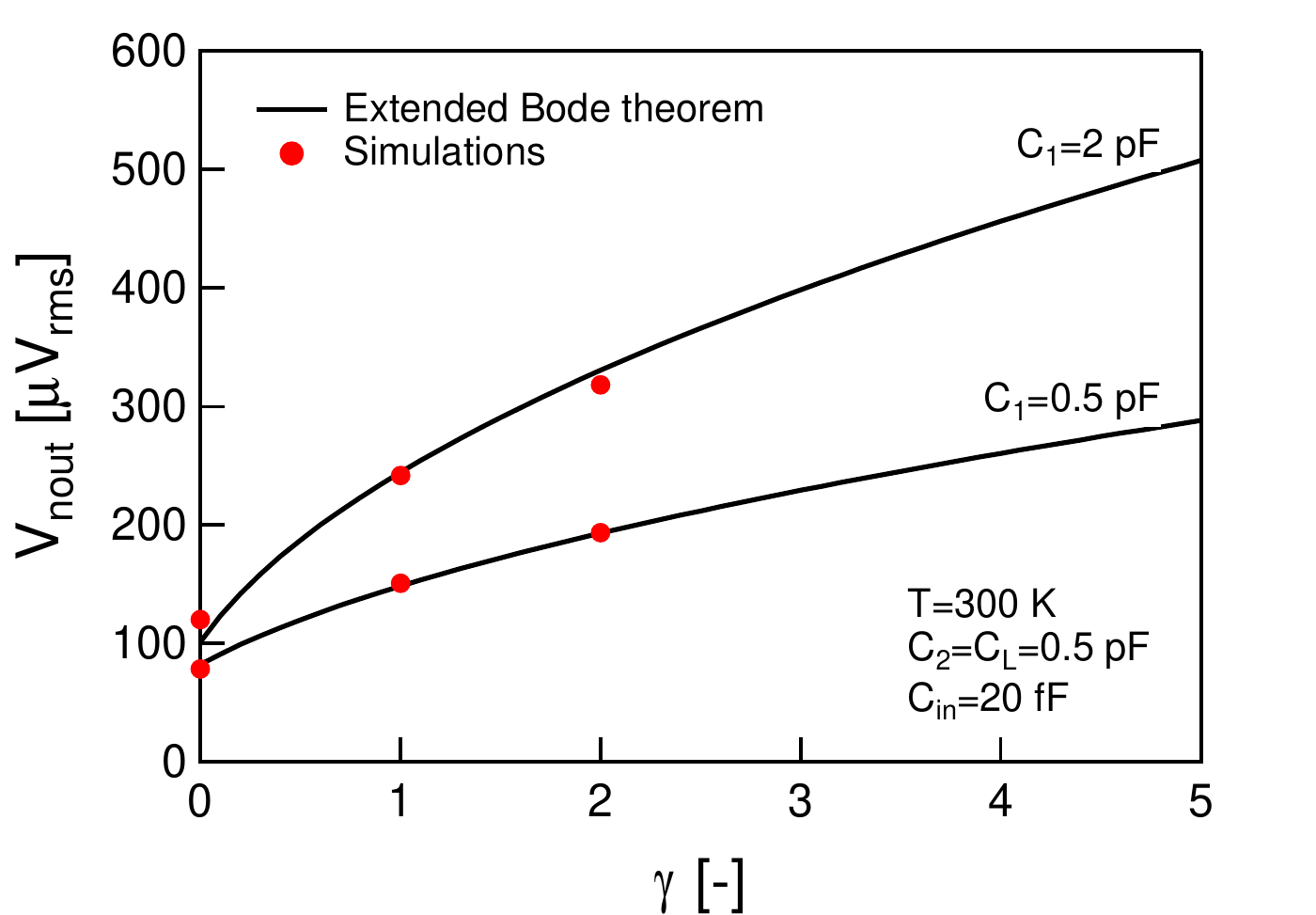}
  \caption{Output noise voltage versus $\gamma$ for $C_1=0.5\;pF$ and $C_1=1\;pF$.}
  \label{fig:SC_amplifier_Vnout_vs_gamma}
\end{figure}

Finally, \cref{fig:SC_amplifier_Vnout_vs_gamma} shows the output noise rms voltage versus the OTA noise excess factor $\gamma$ for two different values of $C_1$ ($C_1=0.5\;pF$ and $C_1=2\;pF$ corresponding to a voltage gain $\aAv=1$ and $\aAv=4$, respectively). As expected from \cref{eqn:SC_amplifier_Vnouttot}, $V_{nout,rms}$ increases as $\sqrt{\bota \cdot \gamma + \bswi}$. The simulation results fall again very close to the value predicted with \cref{eqn:SC_amplifier_Vnouttot}.

\subsection{SC Track \& hold}

\begin{figure}[!ht]
  \centering
  \begin{subfigure}[t]{1\columnwidth}
    \centering
    \includegraphics[width=0.7\columnwidth]{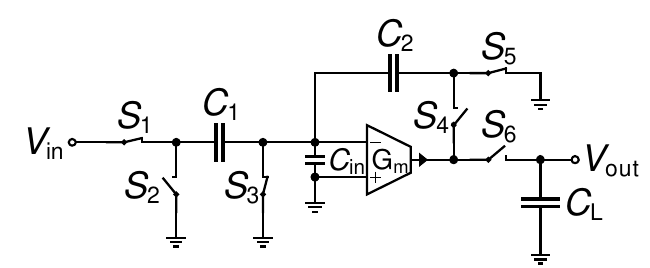}
	\caption{\centering SC track \& hold circuit during \phase{1}.}
	\label{fig:SC_TH_ph1}
  \end{subfigure}%
  \\
  \begin{subfigure}[t]{1\columnwidth}
    \centering
	\includegraphics[width=0.7\columnwidth]{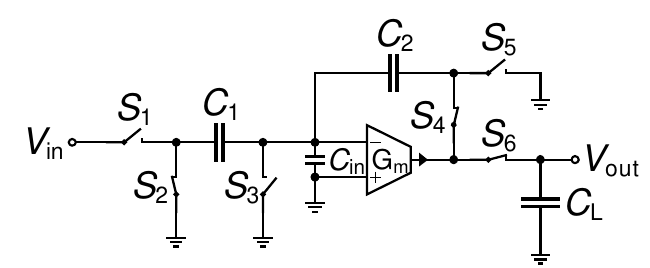}
	\caption{\centering SC track \& hold circuit during \phase{2}}
	\label{fig:SC_TH_ph2}
  \end{subfigure}
  \caption{SC Track \& hold circuit.}
  \label{fig:SC_TH}
\end{figure}

\Cref{fig:SC_TH} shows the schematic of a basic SC track \& hold (TH) circuit which can operate either as a TH or as a SC amplifier featuring a voltage gain set by the capacitance ratio $C_1/C_2$. This circuit operates in two phases as shown in \cref{fig:SC_TH}. During \phase{1}, shown in \cref{fig:SC_TH_ph1}, the sampling capacitor $C_1$ samples the input signal $V_{in}$. During this phase, the feedback capacitor $C_2$ is shorted to be reset, while the capacitor $C_L$ holds the charge that has been sampled at the end of \phase{2} of the previous switching period. During \phase{2}, shown in \cref{fig:SC_TH_ph2}, the charge sampled in $C_1$ is transferred to the feedback capacitor $C_2$. The output voltage seen across $C_L$ is then simply equal to $C_1/C_2 \cdot V_{in}$. This voltage is then sampled and held on $C_L$ at the end of \phase{2}.

\subsubsection{Analysis}
\begin{figure*}[!htb]
  \centering
  \begin{subfigure}[t]{.24\textwidth}
    \centering
	\includegraphics[scale=.6]{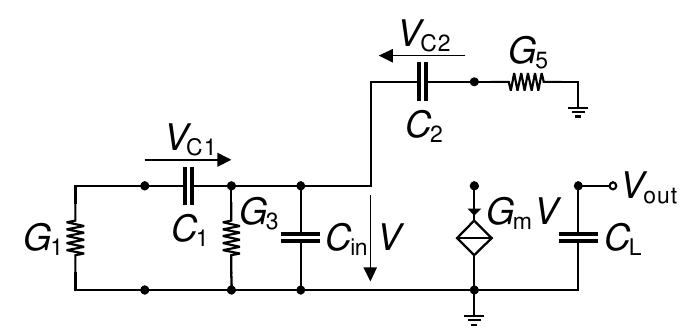}
	\caption{\centering Equivalent SC TH circuit during \phase{1}.}
	\label{fig:SC_TH_ph1a}
  \end{subfigure}
  \begin{subfigure}[t]{.24\textwidth}
    \centering
    \includegraphics[scale=.6]{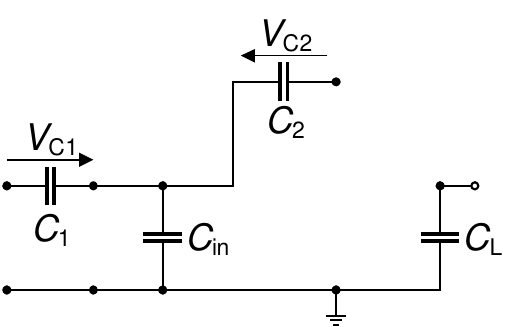}
    \caption{\centering TH circuit for calculation of \Cinf{(kl)}:\\
    $\Cinf{(C_1)}=C_1$, $\Cinf{(C_2)}=C_2$ and $\Cinf{(\Cin)}=\Cin$.}
    \label{fig:SC_TH_ph1b}
  \end{subfigure}
  \begin{subfigure}[t]{.24\textwidth}
    \centering
	\includegraphics[scale=.6]{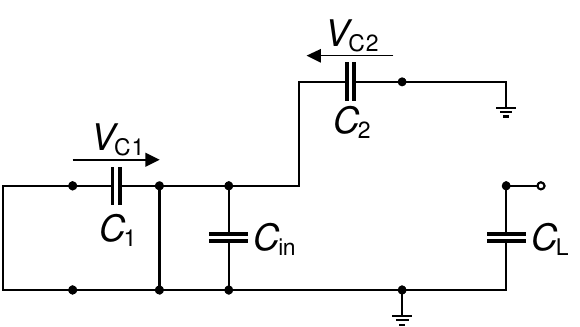}
	\caption{\centering TH circuit for calculation of \Cinfp{(kl)}:\\
    $\Cinfp{(C_1)}=\Cinfp{(C_2)}=\Cinfp{(\Cin)}=\infty$.}
	\label{fig:SC_TH_ph1c}
  \end{subfigure}
  \begin{subfigure}[t]{.24\textwidth}
    \centering
	\includegraphics[scale=.6]{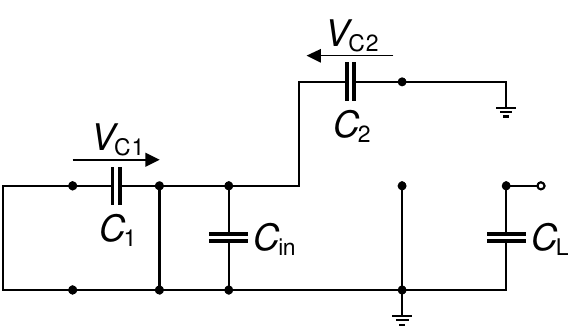}
	\caption{\centering TH circuit for calculation of \Co{(kl)}:\\
    $\Co{(C_1)}=\Co{(C_2)}=\Co{(\Cin)}=\infty$.}
	\label{fig:SC_TH_ph1d}
  \end{subfigure}
  \caption{TH equivalent circuit schematics for \phase{1}.}
  \label{fig:SC_TH_ph1abcd}
\end{figure*}

\begin{figure*}[!htb]
  \centering
  \begin{subfigure}[t]{.24\textwidth}
    \centering
	\includegraphics[scale=.6]{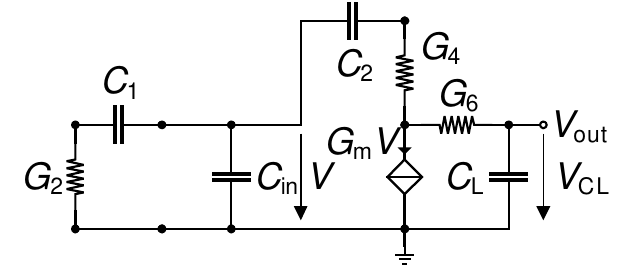}
	\caption{\centering Equivalent SC TH circuit during \phase{2}.}
	\label{fig:SC_TH_ph2a}
  \end{subfigure}
  \begin{subfigure}[t]{.24\textwidth}
    \centering
	\includegraphics[scale=.6]{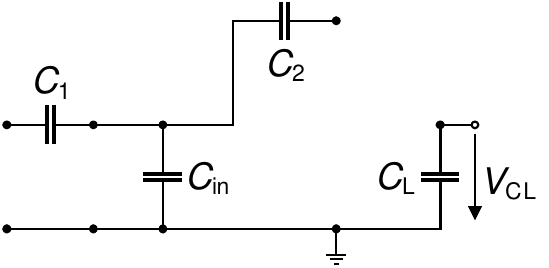}
	\caption{\centering TH circuit for calculation of \Cinf{(kl)}: $\Cinf{(\CL)}=C_L$.}
	\label{fig:SC_TH_ph2b}
  \end{subfigure}
  \begin{subfigure}[t]{.24\textwidth}
    \centering
	\includegraphics[scale=.6]{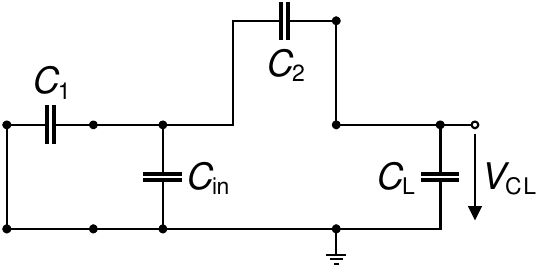}
	\caption{\centering TH circuit for calculation of \Cinfp{(kl)}:\\
    $\Cinfp{(\CL)}=\CL+\frac{C_2 \left(C_1+\Cin \right)}{C_1+C_2+\Cin}$.}
	\label{fig:SC_TH_ph2c}
  \end{subfigure}
  \begin{subfigure}[t]{.24\textwidth}
    \centering
	\includegraphics[scale=.6]{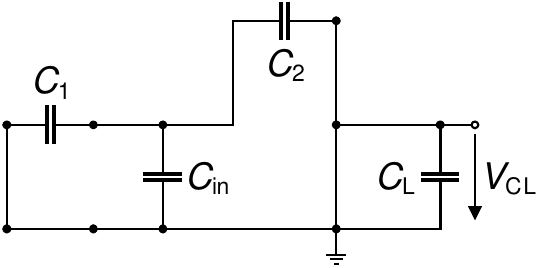}
	\caption{\centering TH circuit for calculation of \Co{(kl)}: $\Co{(\CL)}=\infty$}
	\label{fig:SC_TH_ph2d}
  \end{subfigure}
  \caption{TH equivalent circuit schematics for \phase{2}.}
  \label{fig:SC_TH_ph2abcd}
\end{figure*}

The output voltage is read during \phase{1} from the hold capacitor $C_L$ and the sampled output noise must therefore be calculated at the end of \phase{2}. As in the example above, this circuit presents two non-overlapping phases for which the equivalent linear circuits are depicted in \cref{fig:SC_TH_ph1a} and \cref{fig:SC_TH_ph2a}, respectively.
		
The capacitors sampling a noise charge at the end of \phase{1} are $C_1$, $C_2$ as well as the parasitic capacitance at the OTA input $C_{in}$. The sum of these noise charges is injected into the virtual ground during \phase{2} and transferred to the feedback capacitor $C_2$. This noise charge on capacitor $C_2$ results in a noise voltage at the OTA output which will be sampled on $C_L$ at the end of \phase{2}.

The extended Bode theorem is used to calculate the noise voltage variances across capacitors $C_1$, $C_2$ and $C_{in}$ for \phase{1}. The calculations of capacitors $\Cinf$, $\Cinfp$ and $\Co$ for each capacitor $C_1$, $C_2$ and $\Cin$ during \phase{1} is done using the equivalent circuits shown in \cref{fig:SC_TH_ph1abcd}. The resulting voltage variances based on \cref{eqn:V2n_Bode_OTA} are then given by
\begin{subequations}
  \begin{align}
    \Varp{V}{C_1}{1} &= \kT \cdot \left[\frac{1}{C_1} + 0 - 0 \right] = \frac{\kT}{C_1},\label{eqn:SC_TH_V_C1_ph1}\\
    \Varp{V}{C_2}{1} &= \kT \cdot \left[\frac{1}{C_2} + 0 - 0 \right] = \frac{\kT}{C_2},\label{eqn:SC_TH_V_C2_ph1}\\
    \Varp{V}{\Cin}{1} &= \kT \cdot \left[\frac{1}{\Cin} + 0 - 0 \right] = \frac{\kT}{\Cin}.\label{eqn:SC_TH_V_Cin_ph1}
  \end{align}
\end{subequations}

The total noise charge generated during \phase{1} and injected into the virtual ground is then given by
\begin{equation}\label{eqn:SC_TH_Q_ph1a}
  \begin{split}
    \Varp{Q}{}{1} &=  C_1^2 \cdot \Varp{V}{C_1}{1} + C_2^2 \cdot \Varp{V}{C_2}{1} + \Cin^2 \cdot \Varp{V}{\Cin}{1}\\
                  &= \kT \cdot \left(C_1+C_2+\Cin\right).
  \end{split}
\end{equation}

The charge $\Varp{Q}{}{1}$ is subsequently transferred to capacitor $C_2$ during \phase{2}. Assuming again that the OTA has an infinite DC gain and a zero offset voltage, the output voltage is equal to the voltage across $C_2$ and the variance of the noise voltage seen at the output of the OTA resulting from this charge is hence given by	
\begin{equation}\label{eqn:SC_TH_Q_ph1b}
  \Varp{V}{\CL}{1} = \Varp{V}{C_2}{1} = \frac{\Varp{Q}{}{1}}{C_2^2} = \frac{\kT}{C_2} \cdot \bswip{1}.
\end{equation}
where
\begin{equation}\label{eqn:SC_TH_betaswitch_ph1}
  \bswip{1} = \Av + \ain + 1,
\end{equation}
with $\Av \triangleq C_1/C_2$, $\ain \triangleq \Cin/C_2$.

For \phase{2}, the noise charge held on capacitor $C_L$ can be directly calculated using the extended Bode theorem applied at the output. \cref{fig:SC_TH_ph2abcd} shows the equivalent circuit schematics used for the noise voltage variance calculation using \cref{eqn:V2n_Bode_OTA}, which additionally also requires the feedback gain $\hfb$ given by
\begin{equation}\label{eqn:SC_TH_hfb}
  \hfb = \frac{V}{V_{out}} = \frac{1}{\Av + \ain + 1}.
\end{equation}	
The variance of the noise voltage generated across $C_L$ during \phase{2} is then given by	
\begin{equation}\label{eqn:SC_TH_V_CL_ph2a}
  \Varp{V}{\CL}{2} = \frac{\kT}{C_2} \cdot \left(\gamma \cdot \botap{2} + \bswip{2}\right).
\end{equation}
where
\begin{subequations}
  \begin{align}
    \botap{2} &= \frac{(\Av+\ain+1)^2}{D},\label{eqn:SC_TH_beta_ota}\\
    \bswip{2} &= \frac{1}{\aL} \cdot \frac{\Av+\ain}{D},\label{eqn:SC_TH_beta_sw2}
  \end{align}
\end{subequations}
%\begin{subequations}
%  \begin{align}
%    \botap{2} &= \frac{(1+\Av+\ain)^2}{\Av+\ain+\aL(1+\Av+\ain)},\label{eqn:SC_TH_beta_ota}\\
%    \bswip{2} &= \frac{1}{\aL} \cdot \frac{\Av+\ain}{\Av+\ain+\aL(1+\Av+\ain)},\label{eqn:SC_TH_beta_sw2}
%  \end{align}
%\end{subequations}
where $D$ is given by \cref{eqn:SC_amplifier_D} and $\aL \triangleq \CL/C_2$. The first term in \cref{eqn:SC_TH_V_CL_ph2a} corresponds to the contribution of the OTA during \phase{2} and is actually identical to the expression \cref{eqn:SC_amplifier_betaota_ph2_2} obtained for the SC amplifier. This not surprising since, assuming the inputs are grounded and the switches are ideal (zero resistance and hence noiseless), the circuit of \Cref{fig:SC_TH_ph2} is identical to that of the SC amplifier shown in \Cref{fig:SC_amplifier_ph1}. The second term in \cref{eqn:SC_TH_V_CL_ph2a} corresponds to the contribution of the switches.

In this circuit, none of the capacitors is holding a noise charge from one switching period to the next. Indeed, capacitor $C_2$ is reset during \phase{1}, capacitors $C_1$ and $C_{in}$ are reset during \phase{2} by the action of the OTA, while capacitor $C_L$ is connected to the OTA output during \phase{2} to sample the new value. Consequently, at the end of each switching period, the noise variance of the output voltage corresponds to the sum of the noise injected from phases $\Phi_1$ and $\Phi_2$ without any contributions from the previous switching periods. The variance of the total noise voltage sampled on $C_L$ can hence be expressed as

\begin{equation}\label{eqn:SC_TH_sampled}
  \Var{V}{\CL} = \Varp{V}{\CL}{1} + \Varp{V}{\CL}{2}
  = \frac{\kT}{C_2} \cdot \left(\gamma \cdot \bota + \bswi\right),
\end{equation}
where
\begin{subequations}
  \begin{align}
    \bota &= \botap{2},\label{eqn:SC_TH_beta_ota_tot}\\
    \bswi &= \bswip{1} + \bswip{2},\label{eqn:SC_TH_beta_switch_tot}
  \end{align}
\end{subequations}
The first term in brackets of \cref{eqn:SC_TH_sampled} is the noise contribution coming from the OTA, which only contributes during \phase{2}, while the second term is due to the noise coming from the switches during phases \ph{1} and \ph{2}. Note that \cref{eqn:SC_TH_sampled} matches the result presented in \cite{bib:murmann:sscsmag:2012} except for the second term $\bswip{2}$ in \cref{eqn:SC_TH_beta_switch_tot} which corresponds to the contribution of the switches in \phase{2} and which is omitted in \cite{bib:murmann:sscsmag:2012}. This is reasonable since this term is usually small and can be neglected for the TH circuit because in general $C_{in} \ll C_1=C_2$ ($\Av=1$).
	
\subsubsection{Simulations}
\begin{figure}[!t]
  \centering
  \includegraphics[width=1.0\columnwidth]{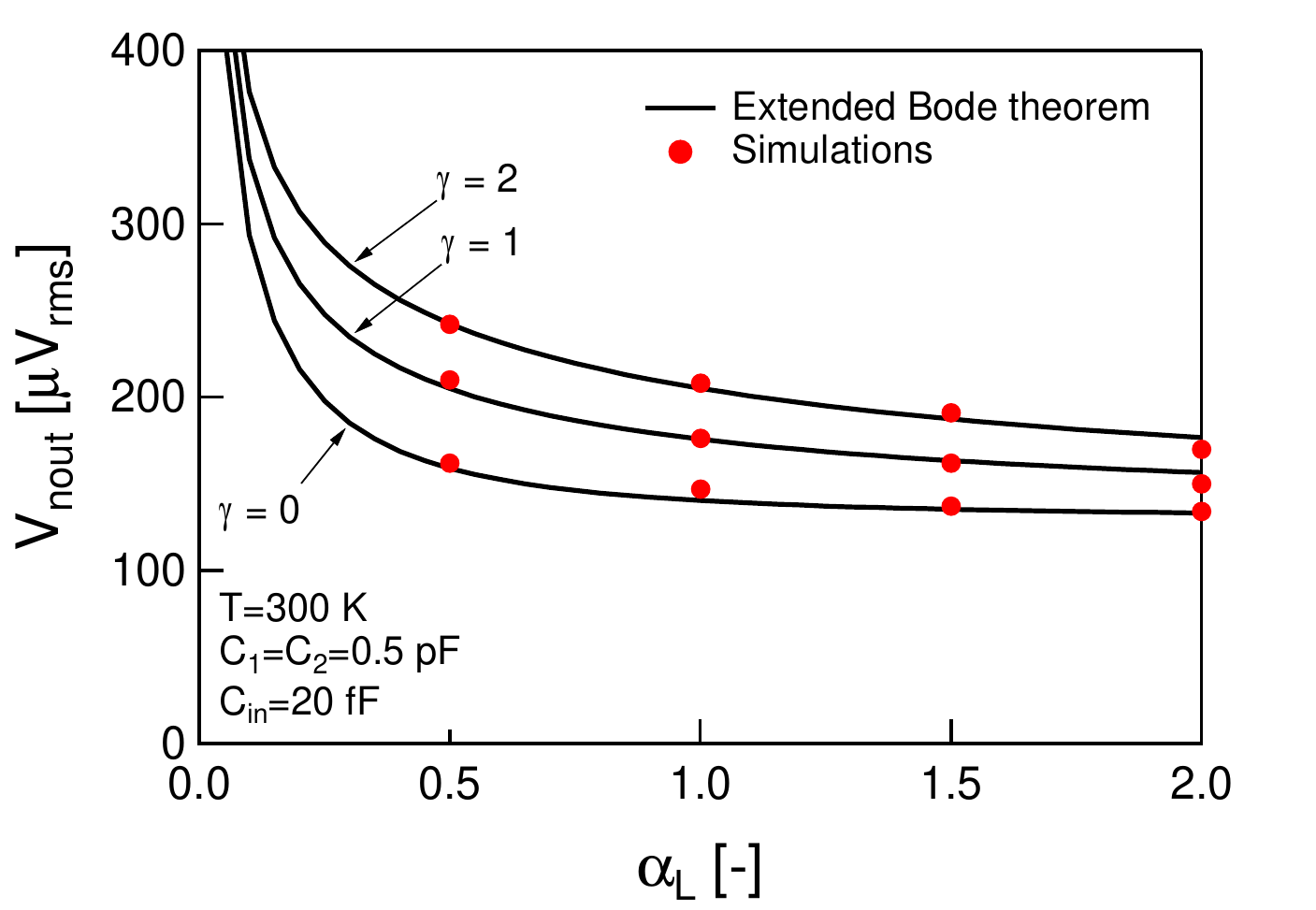}
  \caption{Output noise rms voltage versus $\aL$ for $\gamma=0,1,2$.}
  \label{fig:SC_TH_Vnout_vs_alphaL}
\end{figure}

\begin{figure}[!t]
  \centering
  \includegraphics[width=1.0\columnwidth]{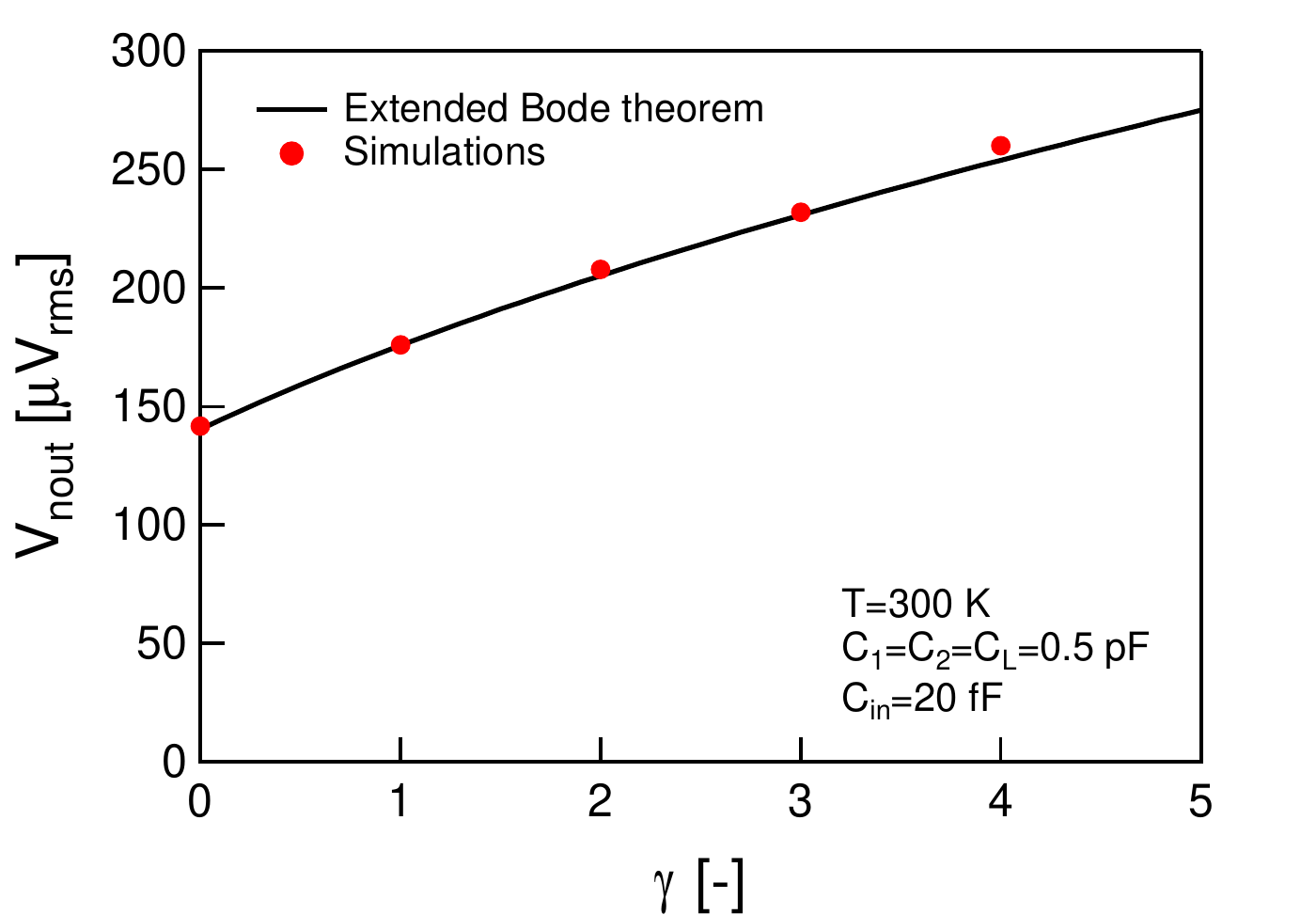}
  \caption{Output noise voltage versus $\gamma$ for $C_1=0.5\;pF$.}
  \label{fig:SC_TH_Vnout_vs_gamma}
\end{figure}

The results obtained for the SC TH have also been validated by transient noise simulation for a gain $\Av=1$ with the same sampling period and temperature as for the SC amplifier. The output noise rms voltage is plotted versus $\aL$ in \Cref{fig:SC_TH_Vnout_vs_alphaL} for 3 different values of $\gamma=0,1,2$. \Cref{fig:SC_TH_Vnout_vs_gamma} shows the output rms noise voltage versus the OTA thermal noise excess factor $\gamma$ for $C_1=C_2=\CL=0.5\;pF$. In both cases, the simulation results are very close to the theoretical results predicted from the extended Bode theorem \cref{eqn:SC_amplifier_Vnouttot}.

\section{Conclusion}\label{sec:conclusion}
The optimization of SC circuits for achieving at the same time low-noise operation at low-power requires an accurate estimation of the integrated noise at the circuit output. Part~I of this paper presents a simple method to obtain an analytical expression of the thermal noise voltage variance at any port of an active SC circuit made of OTAs with a capacitive feedback. The thermal noise variance is derived by simple inspection of three different circuits avoiding the laborious calculation of the noise transfer functions and integrals. It is based on an extension of the original Bode theorem which allows the exact calculation of the thermal noise voltage variance but only in passive circuits \cite{bib:bode:book:1945}. In Part~I, the proposed method is applied to a SC amplifier and a SC \TaH{} circuit and is successfully validated by transient noise simulations. Part~II of the paper will illustrate how this method can be extended to the calculation of thermal noise voltage variances in SC filters.

\section{Appendices}

\subsection{SC Amplifier Noise Calculation}\label{sec:SC_amplifier_appendix}

\begin{figure}[!ht]
  \centering
  \begin{subfigure}[t]{1\columnwidth}
    \centering
    \includegraphics[scale=0.9]{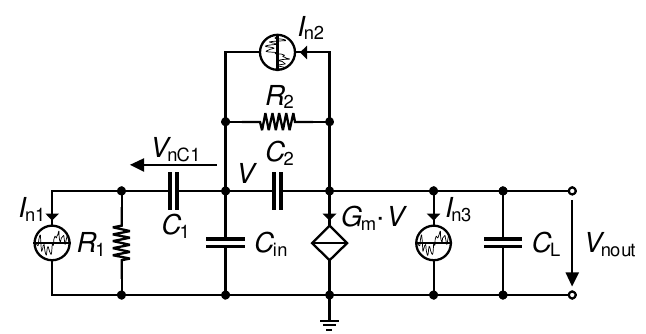}
    \caption{\centering \Phase{1}.}
    \label{fig:SC_amplifier_noise_appendix_ph1}
  \end{subfigure}
  \\
  \begin{subfigure}[t]{1\columnwidth}
    \centering
    \includegraphics[scale=0.9]{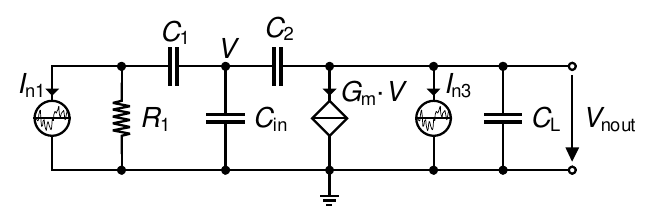}
    \caption{\centering \Phase{2}.}
    \label{fig:SC_amplifier_noise_appendix_ph2}
  \end{subfigure}
  \caption{Small-signal equivalent circuit of \Cref{fig:SC_amplifier} for the calculation of the output noise voltage.}
  \label{fig:SC_amplifier_noise_appendix}
\end{figure}

The noise of the SC amplifier of \Cref{fig:SC_amplifier} can be calculated in a classical way. The noise voltage variance across capacitor $C_1$ during \phase{1} can be calculated from the equivalent small-signal circuit shown in \Cref{fig:SC_amplifier_noise_appendix_ph1} using
\begin{equation}
  \Varp{V}{C_1}{1} = \sum_{i=1}^3 \Varp{V}{C_1,i}{1},
\end{equation}
and
\begin{equation}
  \Varp{V}{C_1,i}{1} = \int_0^{+\infty} |R_{m,i}(f)|^2 \cdot S_{I_{n,i}} \cdot df,
\end{equation}
where $R_{m,i}$ are the noise transfer functions (NTF) (actually transresistances) from the current noise sources $I_{n,i}$ to the voltage across $C_1$ and $S_{I_{n,i}}$ are the PSD of the thermal noise current sources $I_{n,i}$ given by $S_{I_{n,1}}=S_{I_{n,2}}=4\kT/\Ron$ and $S_{I_{n,3}}=4\kT \gamma \Gm$. The NTF $R_{m,i}$ are given by
\begin{equation}
  R_{m,i} \triangleq \frac{V_{nC1}}{I_{n,i}} = R_i \cdot \frac{\num}{\den}
\end{equation}
where the scaling factors $R_i$ and the coefficients of $R_{m,i}$ for $i=1,2,3$ are given in \Cref{tab:SC_amplifier_NTF_terms_ph1}.
For thermal noise, the PSD $S_{I_{n,i}}$ are constant and the noise voltage variance can be calculated using the equivalent noise bandwidth $B_{n,i}$ as
\begin{equation}
  \Varp{V}{C_1,i}{1} = |R_{m,i}(0)|^2 \cdot 4 \kT \cdot S_{I_{n,i}} \cdot B_{n,i}.
\end{equation}
In the case of the SC amplifier in \phase{1}, the NTF are of 3\super{rd}-order. The noise bandwidth and variances can then be obtained by using the expression (22) in \cite{bib:dastgheib:tcas:nov:2008}. The resulting noise voltage variances for each noise source assuming that the switch resistances are negligible ($\Gm \Ron \ll 1$) are given in \Cref{tab:SC_amplifier_V2nC1p1i}. The total noise voltage variance is obtained by summing the 3 contributions, leading to the result shown in the last row of \Cref{tab:SC_amplifier_V2nC1p1i}. Note that even though the individual contributions of the switch noise sources $I_{n,1}$ and $I_{n,2}$ given in \Cref{tab:SC_amplifier_V2nC1p1i} both contain capacitance $C_2$, when summing both contributions, as expected, the result becomes independent of $C_2$ as shown in the fourth row of \Cref{tab:SC_amplifier_V2nC1p1i}.

The same approach can be used to calculate the noise voltage variance at the amplifier output during \phase{2} using the small-signal schematic of \Cref{fig:SC_amplifier_noise_appendix_ph2}. The coefficients of the 2\super{nd}-order NTF are given in \Cref{tab:SC_amplifier_NTF_terms_ph2}. Using the technique presented in \cite{bib:dastgheib:tcas:nov:2008} we get the noise voltage variances at the amplifier output during \phase{2} due to noise sources $I_{n,1}$ and $I_{n,3}$ given in \Cref{tab:SC_amplifier_V2noutp2i}. The total output noise voltage variance is then given by summing these two contributions, leading to the result shown in the last row of \Cref{tab:SC_amplifier_V2noutp2i}.

\newpage

\begin{table*}[!htb]
\centering
\caption{Coefficients of the NTF for the SC-Amplifier in \phase{1}.}
\begin{tabular}{|c|c|c|c|}
  \hline
  Term & $R_{m,1}$ & $R_{m,2}$ & $R_{m,3}$ \\
  \hline
  \hline
  $R_i$ & $\Ron$ & $\frac{1}{\Gm}$ & $-\frac{1}{\Gm}$ \\
  \hline
  $n_2$ & $\frac{\Ron (C_2 \Cin+C_2 \CL+\Cin \CL)}{\Gm}$ & $0$ & $0$ \\
  \hline
  $n_1$ & $\frac{C_2 \Gm \Ron+\Cin+\CL}{\Gm}$ & $\Ron \CL$ & $\Ron C_2$ \\
  \hline
  $n_0$ & $1$ & $0$ & $1$ \\
  \hline
  $d_3$ & \multicolumn{3}{c|}{$\frac{C_1 \Ron^2 (C_2 \Cin+C_2 \CL+\Cin \CL)}{\Gm}$}\\
  \hline
  $d_2$ & \multicolumn{3}{c|}{$\frac{\Ron (C_1 C_2 \Gm \Ron+C_1 C_2+C_1 \Cin+2 C_1 \CL+C_2 \Cin+C_2 \CL+\Cin \CL)}{\Gm}$}\\
  \hline
  $d_1$ & \multicolumn{3}{c|}{$\frac{C_1 \Gm \Ron+C_1+C_2 \Gm \Ron+\Cin+\CL}{\Gm}$}\\
  \hline
  $d_0$ & \multicolumn{3}{c|}{$1$}\\
  \hline
\end{tabular}
\label{tab:SC_amplifier_NTF_terms_ph1}
\end{table*}

\begin{table*}[!htb]
\centering
\caption{Contributions of the various noise sources to $\Var{V}{C_1}$ during \phase{1}.}
\begin{tabular}{|c|c|c|}
  \hline
  Noise source & \multicolumn{2}{c|}{Corresponding noise voltage variance across $C_1$}\\
  \hline
  \hline
  $I_{n,1}$ & $\Var{V}{C_1,1}$ & $\frac{\kT \left(C_1 \left(C_2 (\Cin+\CL)+\Cin^2+3 \Cin \CL+\CL^2\right)+(\Cin+\CL) (C_2
   (\Cin+\CL)+\Cin \CL)\right)}{C_1 (C_1+\Cin+\CL) (C_1 (C_2+\Cin+2 \CL)+C_2
   (\Cin+\CL)+\Cin \CL)}$ \\
  \hline
  $I_{n,2}$ & $\Var{V}{C_1,2}$ & $\frac{\CL^2 \kT}{(C_1+\Cin+\CL) (C_1 (C_2+\Cin+2 \CL)+C_2 (\Cin+\CL)+\Cin
   \CL)}$ \\
  \hline
  $I_{n,3}$ & $\Var{V}{C_1,3}$ & $\frac{\gamma \kT}{C_1+\Cin+\CL}$ \\
  \hline
  Switches only & $\Var{V}{C_1,1}+\Var{V}{C_1,2}$ & $\frac{\kT (\Cin+\CL)}{C_1(C_1+\Cin+\CL)}$ \\
  \hline
  \hline
  Total & $\Var{V}{C_1}$ & $\frac{\kT}{C_1+\Cin+\CL} \cdot \left(\gamma + \frac{\Cin+\CL}{C_1}\right)$ \\
  \hline
\end{tabular}
\label{tab:SC_amplifier_V2nC1p1i}
\end{table*}

\begin{table*}[!htb]
\centering
\caption{Coefficients of the NTF for the SC-Amplifier in \phase{2}.}
\begin{tabular}{|c|c|c|}
  \hline
  Term & $R_{m,1}$ & $R_{m,3}$ \\
  \hline
  \hline
  $R_i$ & $\frac{\Ron C_1}{C_2}$ & $-\frac{C_1+C_2+\Cin}{\Gm C_2}$ \\
  \hline
  $n_1$ & $-\frac{C_2}{\Gm}$ & $\frac{\Ron C_1 (C_2+\Cin)}{C_1+C_2+\Cin}$ \\
  \hline
  $n_0$ & $1$ & $1$ \\
  \hline
  $d_2$ & \multicolumn{2}{c|}{$\frac{\Ron C_1(C_2 \Cin+C_2 \CL+\Cin \CL)}{\Gm C_2}$}\\
  \hline
  $d_1$ & \multicolumn{2}{c|}{$\frac{C_1 C_2+C_2 \Cin+C_1 \CL+C_2 \CL+\Cin \CL+\Gm \Ron C_1 C_2}{\Gm C_2}$}\\
  \hline
  $d_0$ & \multicolumn{2}{c|}{$1$}\\
  \hline
\end{tabular}
\label{tab:SC_amplifier_NTF_terms_ph2}
\end{table*}

\begin{table*}[!htb]
\centering
\caption{Contributions of the various noise sources to $\Var{V}{out}$ during \phase{2}.}
\begin{tabular}{|c|c|c|}
  \hline
  Noise source & \multicolumn{2}{c|}{Corresponding noise voltage variance at the amplifier output}\\
  \hline
  \hline
  $I_{n,1}$ & $\Var{V}{out,1}$ & $\frac{\kT C_1 C_2^2}{(C_2 (\Cin+\CL)+\Cin \CL)(C_1 (C_2+\CL)+C_2
   (\Cin+\CL)+\Cin \CL)}$ \\
  \hline
  $I_{n,3}$ & $\Var{V}{out,3}$ & $\frac{\gamma \kT (C_1+C_2+\Cin)^2}{C_2 (C_1 (C_2+\CL)+C_2 (\Cin+\CL)+\Cin \CL)}$ \\
  \hline
  \hline
  Total & $\Var{V}{out}$ & $\frac{\kT}{C_2} \frac{1}{\Cin \CL+C_1(C_2+\CL)+C_2(\Cin+\CL)}
  \left((C_1+C_2+\Cin)^2 \cdot \gamma + \frac{C_1 C_2^3}{\Cin \CL+C_2(\Cin+\CL)}\right)$ \\
  \hline
\end{tabular}
\label{tab:SC_amplifier_V2noutp2i}
\end{table*}

\clearpage

\bibliographystyle{IEEEtran}

\bibliography{References}

\clearpage

\begin{IEEEbiography}[{\includegraphics[width=1in,height=1.25in,clip,keepaspectratio]{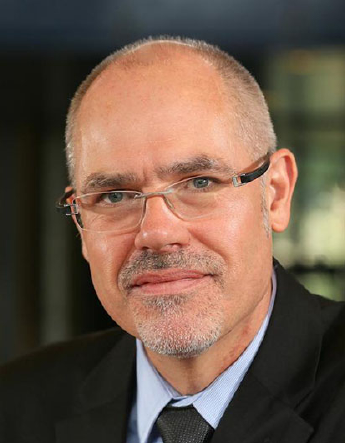}}]{Christian Enz}
(M’84, S'12) received the M.S. and Ph.D. degrees in Electrical Engineering from the EPFL in 1984 and 1989 respectively. He is currently Professor at EPFL, Director of the Institute of Microengineering and head of the IC Lab. Until April 2013 he was VP at the Swiss Center for Electronics and Microtechnology (CSEM) in Neuch\^{a}tel, Switzerland where he was heading the Integrated and Wireless Systems Division. Prior to joining CSEM, he was Principal Senior Engineer at Conexant (formerly Rockwell Semiconductor Systems), Newport Beach, CA, where he was responsible for the modeling and characterization of MOS transistors for RF applications. His technical interests and expertise are in the field of ultralow-power analog and RF IC design, wireless sensor networks and semiconductor device modeling. Together with E. Vittoz and F. Krummenacher he is the developer of the EKV MOS transistor model. He is the author and co-author of more than 250 scientific papers and has contributed to numerous conference presentations and advanced engineering courses. He is an individual member of the Swiss Academy of Engineering Sciences (SATW). He has been an elected member of the IEEE Solid-State Circuits Society (SSCS) AdCom from 2012 to 2014. He is also the Chair of the IEEE SSCS Chapter of Switzerland.
\end{IEEEbiography}

%\vfill
%
\begin{IEEEbiography}[{\includegraphics[width=1in,height=1.25in,clip,keepaspectratio]{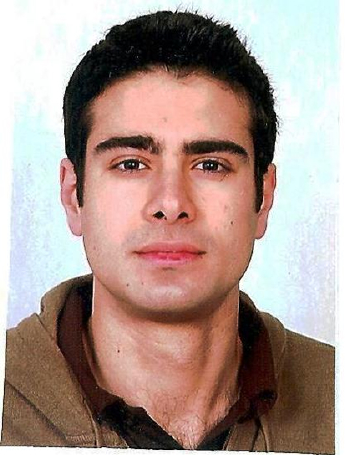}}]{Antonino Caizzone}
Antonino Caizzone was born in Milazzo, Italy, in 1991. He received his bachelor degree in electronic engineering from the University of Catania (Italy) in 2013 and two masters in Micro \& Nano Technologies from INPG Grenoble (France) and Polytechnique of Turin (Italy), respectively, in 2015. He is currently working toward the Ph.D. at the Ecole Polytechnique Federale de Lausanne (EPFL), under the supervision of Prof. Enz and Dr. Boukhayma, on the subject of ultra-low noise and low power sensors for healthcare. Between 2012 and 2013 he worked in STMicroelectronics as an intern on the design of analog electronics on plastic substrate. In 2014, he worked at Georgia Tech. (USA) as visiting researcher on the subject of energy harvesters. .
\end{IEEEbiography}

%\vfill

\begin{IEEEbiography}[{\includegraphics[width=1in,height=1.25in,clip,keepaspectratio]{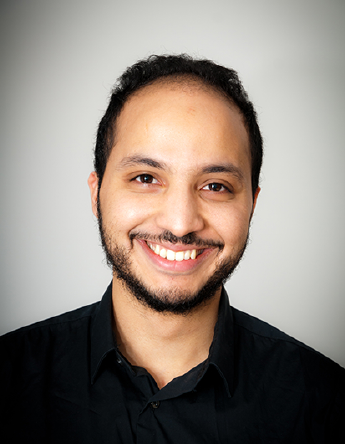}}]{Assim Boukhayma}
received the graduate engineering degree (D.I.) in information and communication technology and the M.Sc. in microelectronics and embedded systems architecture from Institut Mines Telecom (IMT Atlantique), France, in 2013. He was awarded with the graduate research fellowship for doctoral studies from the French atomic energy commission (CEA) and the French ministry of defense (DGA). He received the Ph.D. from EPFL in 2016 on the topic of Ultra Low Noise CMOS Image Sensors. In 2017, he was awarded the Springer Theses prize in recognition of outstanding Ph.D. research. He is currently a scientist at EPFL ICLAB, conducting research in the areas of image sensors and noise in circuits and systems.
From 2012 to 2015, he worked as a researcher at Commissariat a l’Energie Atomique (CEA-LETI), Grenoble, France. From 2011 to 2012, he worked with Bouygues-Telecom as a Telecommunication Radio Junior Engineer.
\end{IEEEbiography}

%\vfill

\begin{IEEEbiography}[{\includegraphics[width=1in,height=1.25in,clip,keepaspectratio]{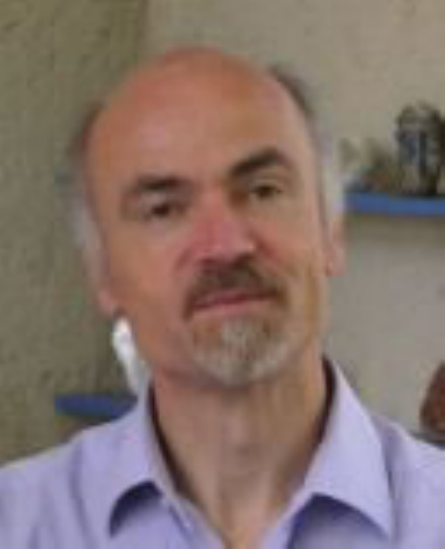}}]{Fran\c{c}ois Krummenacher}
received the M.S. and Ph.D. degrees in electrical engineering from the Swiss Federal Institute of Technology (EPFL) in 1979 and 1985 respectively. He has been with the Electronics Laboratory of EPFL since 1979, working in the field of low-power analog and mixed analog/digital CMOS IC design, as well as in deep sub-micron and high-voltage MOSFET device compact modeling. Dr. Krummenacher is the author or co-author of more than 120 scientific publications in these fields. Since 1989 he has also been working as an independent consultant, providing scientific and technical expertise in IC design to numerous local and international industries and research labs.
\end{IEEEbiography}

\end{document}